\documentclass[twocolumn]{aastex6}

\usepackage{graphicx}
\usepackage{amsmath}

\begin{document}


\title{Tracing the Dynamical Mass in Galaxy Disks Using HI Velocity Dispersion and its Implications for the Dark Matter Distribution in Galaxies}



\author{Mousumi Das\altaffilmark{1}}
\affil{Indian Institute of Astrophysics, Koramangala, Bangalore, Karnataka 560034, India}
\author{Stacy S. McGaugh}
\affil{Department of Astronomy, Case Western Reserve University, 10900 Euclid Ave, Cleveland, OH 44106, USA}
\author{Roger Ianjamasimanana}
\affil{Department of Physics and Electronics, Rhodes University, PO Box 94, Grahamstown 6140, South Africa}
\and
\affil{South African Radio Astronomy Observatory, 2 Fir Street, Black River Park, Observatory 7925, Cape Town, South Africa}
\author{James Schombert }
\affil{Department of Physics, University of Oregon, 120 Willamette Hall,  1371 E 13th Avenue, Eugene, OR 97403461, USA}
\and
\author{K.S.Dwarakanath}
\affil{Astronomy and Astrophysics, Raman Research Institute, C. V. Raman Avenue, 5th Cross Road, Sadashivanagar, Bengaluru, Karnataka 560080, India}




\altaffiltext{1}{email~:~chandaniket@gmail.com, mousumi@iiap.res.in}

\begin{abstract}
We present a method to derive the dynamical mass of face-on galaxy disks using their neutral hydrogen (H\,{\sc i}) velocity dispersion ($\sigma_{\rm{H\,{\textsc i}}}$). We have applied the method to nearby, gas rich galaxies that have extended H\,{\sc i} gas disks and have low inclinations. The galaxy sample includes 4 large disk galaxies; NGC628, NGC6496, NGC3184, NGC4214 and 3 dwarf galaxies DDO46, DDO63 and DDO187. We have used archival H\,{\sc i} data from the {\it THINGS} and {\it LITTLE THINGS} surveys to derive the H\,{\sc i} gas distributions and {\it SPITZER} mid-infrared images to determine the stellar disk mass distributions. We examine the disk dynamical and baryonic mass ratios in the extreme outer disks where there is H\,{\sc i} gas but no visible stellar disk. We find that for the large galaxies the disk dynamical and H\,{\sc i} gas mass surface densities  are comparable in the outer disks. But in the smaller dwarf galaxies, for which the total H\,{\sc i} gas mass dominates the baryonic mass i.e. M(HI)$\geq$M(stars), the disk dynamical mass is much larger than the baryonic mass. For these galaxies there must either be a very low luminosity stellar disk which provides the vertical support for the H\,{\sc i} gas disk or there is halo dark matter associated with their disks, which is possible if the halo has an oblate shape so that the inner part of the dark matter halo is concentrated around the disk. Our results are important for explaining the equilibrium of H\,{\sc i} disks in the absence of stellar disks, and is especially important for gas rich, dwarf galaxies that appear to have significant dark matter masses associated with their disks. 
\end{abstract} 




\section{Introduction} 

Neutral hydrogen (HI) gas is one of the best tracers of the dynamical mass distributions in galaxies \citep{deblok.etal.2008,oh.etal.2015}. This is because of its cold, dissipational nature but also because it is the most extended, visible component of disk galaxies with radii typically twice the size of the stellar disks \citep{bigiel.etal.2010,swaters.etal.2002,mishra.etal.2017}. This is especially true for late type galaxies and low luminosity dwarfs where the H\,{\sc i} distribution is usually several times the radius of the visible disk \citep{swaters.etal.2002,hunter.etal.2012}. This makes H\,{\sc i} an excellent tracer of galaxy rotation and the dynamical masses of galaxies \citep{bosma.1981,vanalbada.etal.1985,begeman.1989}. Hence, it is also a good tracer of the dark matter distribution in galaxies \citep{katz.etal.2019}. 

The H\,{\sc i} distribution and kinematics can also be used to constrain the halo density profiles of galaxies using their inner rotation curves \citep{deblok.2010,gentile.etal.2004,vandenbosch.etal.2000}. In recent times this has become an important test of the cold dark matter (CDM) models of galaxy formation which predict cuspy halo profiles rather than isothermal core profiles (cusp-core problem) \citep{kuzio.etal.2008,deblok.2010}. The ideal galaxies for such studies are low surface brightness (LSB) dwarf galaxies that are dark matter dominated even in their inner disks \citep{mcgaugh.deblok.1998}. Studies of LSB galaxies have shown that they have the largest fraction of dark matter in our local Universe \citep{deblok.etal.1997} and their H\,{\sc i} mass typically forms the major fraction of their baryonic mass \citep{mcgaugh.etal.2000}.  

Apart from measuring disk rotation, the H\,{\sc i} disks in edge on galaxies have been used to constrain the large scale shape of dark matter halos 
\citep{peters.etal.2017,banerjee.etal.2010,olling.1996,newone.1995}. Warped H\,{\sc i} disks have been used to investigate the alignment of the disk and halo angular momenta through modeling of the flared H\,{\sc i} disks \citep{debattista.sellwood.1999,haan.braun.2014}. In all these studies, the vertical gravitational force binding the H\,{\sc i} gas is assumed to be solely due to the stellar disk \citep{patra.etal.2014,malhotra.1995}. This assumption has also been used to model the vertical H\,{\sc i} distribution and hence constrain the stellar disk height or vertical density profile \citep{degrijs.peletier.1997}. 

In very general terms, the H\,{\sc i} disk rotation is supported by the gravitational force of the galaxy and especially the dark matter halo. The vertical support of the H\,{\sc i} disk is due to the stellar disk potential which provides the gravity to bind the gas disk. Gravity tries to collapse the disk but is balanced by the velocity dispersion of the gas. 
Several mechanisms have been invoked to provide the energy input for the  H\,{\sc i} velocity dispersion, such as active galactic nuclei (AGN) activity or stellar winds that increase the H\,{\sc i} velocity dispersion and turbulent gas pressure  \citep{tamburro.etal.2009,krumholz.etal.2016}. However, the extended H\,{\sc i} disks of galaxies such as LSB dwarfs and extreme late type spirals are not generally associated with AGN activity nor do they have ongoing massive star formation \citep{das.etal.2019,das.etal.2007} and so it is not clear what mechanism feeds the vertical dispersion of gas in such galaxies \citep{stilp.etal.2013}.

A more puzzling problem is that galaxies with extended H\,{\sc i} disks have hardly any visible disk mass in their outer regions. But a stellar disk is essential to provide the gravitational support for the vertical gas distribution, especially for H\,{\sc i} rich dwarf galaxies and LSB galaxies  \citep{kreckel.etal.2011,mishra.etal.2017}, where even deep near-infrared (NIR) observations detect only very diffuse or low surface density stellar disks \citep{honey.etal.2016}. For such galaxies the stellar disk may not be massive enough to provide the vertical gravity for the extended H\,{\sc i} disk and so the assumption that the H\,{\sc i} disk is gravitationally supported by a stellar disk may not always hold. One of the ways to solve this problem is to include the presence of dark matter in galaxy disks so that it can help support the vertical equilibrium of the HI dominated disks. The existence of disk dark matter was first postulated by Oort (Oort 1932) and has recently been studied by several authors as a component of the Galactic disk \citep{kramer.randall.2016,fan.etal.2013,saburova.etal.2013,kalberla.etal.2007}. The disk dark matter could be part of an oblate halo or it could be mass that has settled within the disk as a result of mergers with satellite galaxies \citep{read.etal.2008}. 

In this study we investigate the dynamical mass of the outer disks of galaxies using H\,{\sc i} velocity dispersion $\sigma_{\rm{H\,{\textsc i}}}$ as a tracer of the local vertical disk potential. In the past few years, high quality H\,{\sc i} observations have made it possible to trace the radial distributions of $\sigma_{\rm{H\,{\textsc i}}}$ for a number of nearby galaxies \citep{ianjamasimanana.etal.2015,mogotsi.etal.2016,ianjamasimanana.etal.2017} as well as probe the cold and warm components of gas in nearby galaxies \citep{mogotsi.etal.2016,ianja.etal.2012}. These studies show that large column densities of H\,{\sc i} exist in regions where there is hardly any stellar disk mass, good examples being the gas rich dwarfs DDO63 and DDO46 in the {\it LITTLE THINGS} survey \citep{hunter.etal.2012}. 

In the following sections we first derive an expression for the surface density of the disk dynamical mass assuming that the vertical gravitational support of the H\,{\sc i} gas disk is solely due to the stellar and gas masses. We then use H\,{\sc i} data from the {\it THINGS} and {\it LITTLE THINGS} surveys to determine the dynamical disk mass surface densities in 7 nearby face-on disk galaxies and compare it with their baryonic mass surface densities using {\it SPITZER IRAC} 3.6$\rm{\mu}$m images. Finally, we compare the dynamical and H\,{\sc i} masses in the outer disks and discuss the implications of our results for understanding the dark matter distribution in galaxies. 
 
\section{The Disk Dynamical Mass}

In this section we derive an expression for the vertical equilibrium of a disk assuming only gravitational support from stars and gas. Our treatment is mainly applicable to the outer regions of gas rich disk galaxies. We use H\,{\sc i} as a tracer of the vertical potential so that the disk dynamical mass can be estimated using the H\,{\sc i} velocity dispersion. There are two main conditions for applying this method. (i)~There should be very little star formation and no nuclear activity in the regions where the dynamical mass is calculated as such activity will increase the turbulent gas velocity ($v_{turb}$) and the equations do not include $v_{turb}$. (ii)~The disk should be close to face on i.e. have low inclination. (iii)~We neglect the molecular hydrogen ($H_2$) gas mass. This is not a bad assumption for the outer regions of galaxy disks which contain little or no molecular gas \citep{dessauges-zavadsky.etal.2014} or for LSB dwarf galaxies where CO is not detected \citep{das.etal.2006}.  

\subsection{Derivation of the Disk Dyamical Mass surface density $\Sigma_{dyn}$}

We start with the Jeans and Poisson equations for an axisymmetric system (Binney \& Tremaine 1988). Assuming that the system is time independent and the velocities along the r, $\phi$, z directions are independent, we have $-\rho_{i}\frac{d\Phi}{dz}=\frac{d[\rho_{i}{\sigma_{z}}^{2}]}{dz}$, where $\rho_{i}$ represents the tracer density, and $\sigma_z$ is the velocity dispersion in the vertical direction. Using the Poissons equation and assuming H\,{\sc i} to be the tracer of the potential, we obtain the following equations.

\begin{equation}
\frac{1}{\rho_{HI}}\frac{d[{\sigma_{\rm{zH\,{\textsc i}}}}^{2}\rho_{HI}]}{dz}=-\frac{d\Phi}{dz}
\end{equation}
\begin{equation}
\frac{d^{2}\Phi}{dz^{2}}=4\pi G[\rho_{d}~+~\rho_{HI}] 
\end{equation}

where $\rho_{d}$ is the disk stellar mass density, $\rho_{HI}$ is the H\,{\sc i} gas density and $\sigma_{\rm{zH\,{\textsc i}}}$ is the H\,{\sc i} disk velocity dispersion in the vertical or z direction. Here we have assumed that the H\,{\sc i} gas is a static, smooth atmosphere in equilibrium with the disk potential. The force in the vertical direction is given by,

\begin{equation}
K_{z}(z)=-4\pi~G\int_{0}^{z}~[\rho_{d}+\rho_{HI}]~dz
\end{equation} 

If we assume that the stellar and gas disks have exponential mass distributions given by $\rho_{d}=\rho_{e}~e^{-z/z_{e}}$ and $\rho_{HI}=\rho_{g0}~e^{-z/z_{g0}}$, \citep{vanderkruit.1988}, where $z$ is the distance from the $z=0$ plane then,

\begin{equation}
\begin{split}
\frac{1}{\rho_{HI}}\frac{d[{\sigma_{\rm{zH\,{\textsc i}}}}^{2}\rho_{HI}]}{dz}=-4\pi Gz_{e}\rho_{e}(1-e^{-z/z_{e}})\\
~-~4\pi Gz_{g0}\rho_{g0}(1-e^{-z/z_{g0}})
\end{split}
\end{equation}

We have used an exponential form for the vertical mass distribution in the stellar disk and gas disk because it is the simplest form.
Integrating both sides of the equation, from z=0 to $\infty$, we obtain,

\begin{equation}
\begin{split}
\int_{0}^{\infty}d[{\sigma_{\rm{zH\,{\textsc i}}}}^{2}\rho_{HI}]=-4\pi Gz_{e}\rho_{e}\int_{0}^{\infty}(1-e^{-z/z_{e}})\rho_{HI}dz \\
~-~4\pi Gz_{g0}\rho_{g0}\int_{0}^{\infty}(1-e^{-z/z_{g0}})\rho_{HI}dz 
\end{split}  
\end{equation}

The first integration on the right yields the term -$4\pi Gz_{e}\rho_{e}\rho_{g0}(z_{g0} - \frac{z_{g0}z_{e}}{(z_{g0} + z_{e})})$ while the second gives -$2\pi G{{z_{g0}}^{2}}{\rho_{g0}}^{2}$ (where integration of the type $\int_{0}^{\infty}e^{-\frac{z}{z_e}}dz=z_{e}$ and $\int_{0}^{\infty}e^{-\frac{2z}{z_{g0}}}dz=z_{g0}/2$). For the left hand side we can assume that the gas density vanishes at large z. Then we obtain the following equation,

\begin{equation}
\frac{\sigma_{\rm{zH\,{\textsc i}}}^{2}}{2\pi Gz_{g0}}=2\frac{\rho_{e}z_{e}z_{g0}}{(z_{e}+z_{g0})}~+~\rho_{g0}z_{g0} 
\end{equation}

The total disk surface density $\Sigma_{d}$ including the stellar and gas components is given by the relation,

\begin{equation}
\Sigma_{d}=2\int_{0}^{\infty}(\rho_{d}~+~\rho_{HI})~dz=2z_{e}\rho_{e}~+~2z_{g0}\rho_{g0}
\end{equation}

The z component of the H\,{\sc i} velocity dispersion $\sigma_{\rm{zH\,{\textsc i}}}$ is hence related to the total stellar and gas mass disk surface densities, $\Sigma_{s}$ and $\Sigma_{\rm{zH\,{\textsc i}}}$, in the following way.

\begin{equation}
\frac{{\sigma_{\rm{zH\,{\textsc i}}}}^{2}}{\pi~Gz_{g0}}=\frac{2z_{g0}}{(z_{e}+z_{g0})}\Sigma_{s}~+~\Sigma_{HI}
\end{equation} 

This relation can also be written as, 

\begin{equation}
\frac{{\sigma_{zHI}}^{2}}{\pi~Gz_{g0}}=\Sigma_{s}~+~\Sigma_{HI} + \Sigma_{s}\frac{1-\frac{z_{e}}{z_{g0}}}{1+\frac{z_{e}}{z_{g0}}}
\end{equation} 

Thus the disk dynamical mass traced by $\sigma_{\rm{zH\,{\textsc i}}}$ is composed of two parts. The first is the sum of the stellar and gas mass surface densities, whereas the second is a coupling term that is due to the difference in the scale heights of the stellar and gas disks. For example if $z_{e}=z_{g0}$, then, the third term in equation~[9] vanishes and we have,

\begin{equation}
\frac{{\sigma_{\rm{zH\,{\textsc i}}}}^{2}}{\pi~Gz_{e}}=\Sigma_{s}~+~\Sigma_{HI}
\end{equation} 
  
Until now we have assumed that the vertical support is only due to the visible stellar disk and the H\,{\sc i} gas mass. For the outer disks of galaxies where there is no visible stellar disk $\Sigma_{s}\sim 0$. Then the left hand side then represents the dynamical mass surface density of the disk and is given by,

\begin{equation}
\Sigma_{dyn}=\frac{{\sigma_{zHI}}^{2}}{\pi Gz_{g0}} 
\end{equation} 

The mean H\,{\sc i} velocity dispersion in the z direction $\sigma_{\rm{zH\,{\textsc i}}}$ can be derived from the width of the H\,{\sc i} velocity profiles of face-on galaxies \citep{petric.rupen.2007,ianjamasimanana.etal.2015}. Then using equation~[9] we can determine the dynamical mass in the disk ($\Sigma_{dyn}$). The value of $\Sigma_{dyn}$ can be compared with the stellar and gas surface densities ($\Sigma_{s}~+~\Sigma_{HI} + \Sigma_{s}\frac{1-\frac{z_{e}}{z_{g0}}}{1+\frac{z_{e}}{z_{g0}}}$), which can be determined from the near-infrared images and H\,{\sc i} intensity maps. 

This comparison is especially important for the outer disks of galaxies where there is no visible stellar disk supporting the H\,{\sc i} disk. In these regions the equilibrium is given simply by taking $\Sigma_{s}\approx 0$, so that we get, $\Sigma_{dyn}=\Sigma_{HI}$. It must be noted that in such cases when there is no visible stellar disk, the value of $z_{g0}$ will correspond to the scale height of the dynamical mass component that provides the binding gravity for the H\,{\sc i} gas disk. This is important for understanding the nature of the dynamical mass component, whether it is part of an oblate halo or whether it is dark matter within the disk as discussed in Section~5.  

However, for disk regions where there is no visible stellar disk and $\Sigma_{dyn}>\Sigma_{HI}$, it means that the disk dynamical mass is larger than the detected baryonic mass of the disk. This could indicate the presence of a very low luminosity, undetected stellar disk; alternatively there could be halo dark matter asssociated with the disk in the form of an oblate inner halo or a thick disk. The mass surface density of the dark matter associated with the disk $\Sigma_{dm}$, can be determined from the difference between the dynamical and H\,{\sc i} disk i.e.

\begin{equation}
\Sigma_{dm}=\Sigma_{dyn}-\Sigma_{HI}  
\end{equation}

In the following sections we apply this method to the outer H\,{\sc i} disks of 4 large galaxies that are close to face-on and 3 dwarf galaxies that are also close to face-on. Our main aim is to see if $\Sigma_{dyn}$ is larger than $\Sigma_{HI}$ in the outer regions of galaxy disks and if so, by how much.

\subsection{Velocity Dispersion in a Near Face-on Disk}   

This method can be applied only to face-on galaxies that have an inclination angle ($i$) close to 0. This is to minimize the rotational velocity component along the line of sight so that the velocity gradient is smaller within the finite beam of the telescope. It allows a more reliable estimate of $\sigma_{z\rm{H\,{\textsc i}}}$ from the width of the H\,{\sc i} line profiles \citep{vanderkruit.shostak.1982}. 
But such galaxies are hard to find, especially if high resolution momemt maps of the H\,{\sc i} distribution are also essential. Hence, we have relaxed the criterion to include galaxies with $i<$40$^{\circ}$ (Table~1). The value of $\sigma_{\rm{zH\,{\textsc i}}}$ can then be approximately determined by averaging the values over concentric annuli in the disks. If we assume that $\frac{\sigma_{\rm{zH\,{\textsc i}}}}{\sigma_{r}}=0.5$ \citep{dehnen.binney.1998}, and using cylindrical coordinates (r,$\phi$,z) for the disk, then from  \citet{das.etal.2019},

\begin{equation}
{\sigma_{\rm{zH\,{\textsc i}}}}^{2}=\frac{{\sigma_{\rm{H\,{\textsc i}}}}^{2}}{cos^{2}i~+~4sin^{2}i}   
\end{equation}

where $\sigma_{\rm{H\,{\textsc i}}}$ is derived from the width of the H\,{\sc i} profiles. The correction factor introduced by equation[13] ranges in value from 1 (at 0 degrees) to 0.45 (at 45 degrees). It leads to a lower estimate of the disk dynamical mass, so not using the correction would lead to higher dynamical mass estimates.

\subsection{The Scale Height of the H\,{\sc i} Gas Disk}

Observations of edge on disk galaxies show that the cool H\,{\sc i} disk can have a FWHM of 260~pc in the central region of the disk but may flare out to a FWHM of 1.6~kpc in the extreme outer parts \citep{zschaechner.rand.2015,matthews.wood.2003}. In the H\,{\sc i} plots of \citet{matthews.wood.2003} we find that to a first approximation the H\,{\sc i} disk thickness appears to be fairly uniform and has a value of $\approx\pm 30^{\prime\prime}$ or 0.75kpc. The extended disks in these observations are similar to the H\,{\sc i} dominated outer disks of late type galaxies and the diffuse disks of gas rich dwarfs. Other observations of edge on disk galaxies indicate that the H\,{\sc i} disks have a $z_{1/2}$(H\,{\sc i}) value of $\approx$0.5~kpc as shown in Figure~25 of \citet{obrien.etal.2010}, unless there is a warping or flaring of the outer disk. 

In our calculations the $z_{1/2}$(H\,{\sc i}) value represents the disk scale height that contains most of the H\,{\sc i} mass. Hence we have assumed the disk thickness corresponds to the density $0.1\rho_{g0}$, where $\rho_{HI}=\rho_{g0}~e^{-z/z_{g0}}$. This height contains 90\% of the H\,{\sc i} surface density as shown below.

\begin{equation}
\Sigma_{HI}(<z_{g0})=2\int_{0}^{ln10}\rho_{g0}e^{z/z_{g0}}dz~=~0.9\times 2\rho_{g0}z_{g0}
\end{equation}
 
Hence, the exponential scale height of the vertical H\,{\sc i} gas distribution $z_{g0}$ is related to the half z height $z_{1/2}$ by,

\begin{equation}
z_{1/2}~=~ln(10)z_{g0}~=~2.3026z_{g0}
\end{equation}

An accurate estimate of $\Sigma_{dyn}$ should include a radial variation of $z_{1/2}$ for both the stellar and H\,{\sc i} disks as they both affect the values of $\Sigma_{dyn}$. But since this study is mainly aimed at comparing the $\Sigma_{dyn}$ with the $\Sigma_{\rm{H\,{\textsc i}}}$ in the outer disks of galaxies, we will use constant dynamical disk width values of $z_{1/2}$=0.5~kpc for the large disk galaxies (NGC628, NGC6946, NGC3184 and NGC4214). For the dwarf galaxies, since they  have puffier stellar/H\,{\sc i} disks (DDO46, DDO63 and DDO187) we will use two values of $z_{1/2}$=0.5 and 1~kpc. The stellar disk thickness is relevant for only one galaxy (DDO187) as discussed in section 4. {We will assume a similar relation as equation[15] for the stellar exponential vertical scalelength $z_{e}$.}

\begin{figure*}
\centering{
\includegraphics[scale=0.65,trim=0 1cm 0 0,clip]{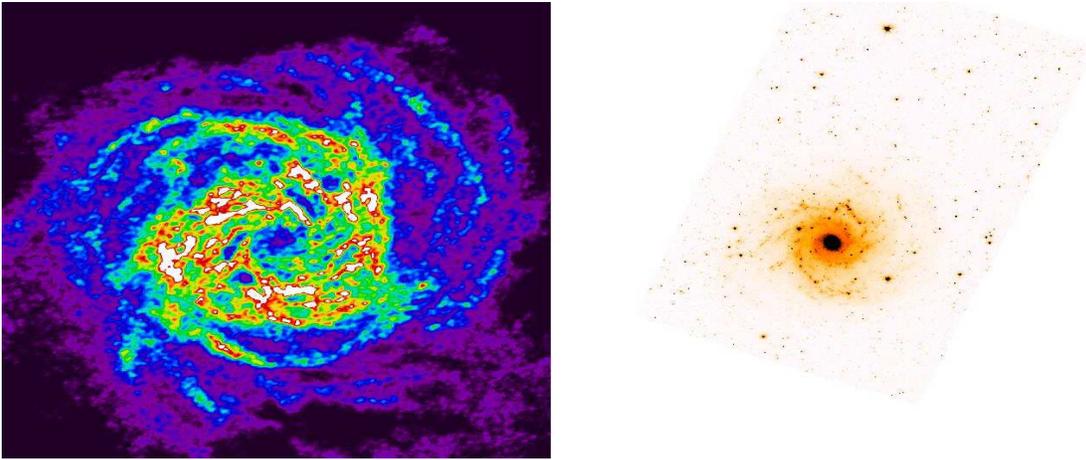}
\caption{(a)~The panel above shows images of NGC628 at two different wavelengths, all on the same scale. The left panel is the moment0 H\,{\sc i} image from the {\it THINGS} survey and the right hand side panel is the IRAC 3.6$\rm{\mu}$m image. Note that the H\,{\sc i} extends out to more than twice the stellar disk extent.}}  
\end{figure*} 

\section{Applying our method to Nearby Face-on Galaxies}

\subsection{Sample Selection}

We have chosen our galaxies based on the following criterion :\\
(i)~The H\,{\sc i} gas should extend well beyond their star forming stellar disks. This is because the expression in equation~[9] holds for H\,{\sc i} disks in hydrostatic equilbrium and this may not apply to regions where there are AGN outflows or massive star forming regions which can increase the turbulent gas pressure and destroy the disk equlibrium. However, on average over sections of the disk, the mechanical energy input from star formation may still produce a H\,{\sc i} gas layer in dynamical equilibrium with the gravitational potential. So our assumption is that whatever causes the H\,{\sc i} velocity dispersion, the gas is largely in equilibrium from a dynamical stand point.\\
(ii)~The galaxies should be nearby so as to obtain several beam widths across the disks and hence several values of the azimuthally averaged H\,{\sc i} velocity dispersion. \\
(iii)~The galaxies should have low inclinations in order to trace the vertical motion (z direction). We have finally selected seven galaxies (Table~1) that have relatively high H\,{\sc i} gas masses, are at distances 2 to 11~Mpc and have inclinations less than 40$^{\circ}$. 

The first four galaxies NGC628, NGC6946, NGC3184 and NGC4214 are large, late type spirals that have H\,{\sc i} to stellar disk mass ratios M(HI)/M(star)$<$1 (Table~1) and host star formation in their inner regions. But in all of them the H\,{\sc i} gas disk extends well beyond the $R_{25}$ radius of the galaxy and the old stellar disk. They are all part of the {\it The H\,{\sc i} Nearby Galaxy Survey ({\it THINGS})} survey \citep{walter.etal.2008}. The galaxies NGC628, NGC6946 and NGC3184 are grand design spirals with moderate star formation rates (SFRs) of values 1 to 5, but NGC4214 has a much lower SFR of 0.05 \citep{walter.etal.2008}. The size, gas rich nature and SFR of NGC4214 is more akin to that of late type dwarf galaxies \citep{das.etal.2019} than large spirals. The galaxies NGC628 and NGC6946 also have compact H\,{\sc ii} regions in their outer disks \citep{ferguson.etal.1998,lelievre.Roy.2000}, which is rare and often attributed to cold gas accretion from the intergalactic medium which can trigger star formation
\citep{thilker.etal.2007}. Their outermost star forming regions are associated with their inner spiral arms and can be traced into the outer disk using {\it GALEX} UV images \citep{gildepaz.etal.2007}. 

Figure~1 shows an example of how the H\,{\sc i} disk can extend well beyond the stellar disks of late type galaxies such as NGC628. The H\,{\sc i} disk is about three times the size of the stellar disk. This is a well known type~I extended UV (XUV) galaxy \citep{thilker.etal.2007}; such galaxies are characterised by UV emission from star formation along spiral arms in their extended disks. Most of the UV emission coincides with the stellar disk but faint trails of UV emission extend into the outer H\,{\sc i} disk, following the disk spiral structure. 

The remaining 3 galaxies, DDO46, DDO63 and DDO187 are gas rich dwarfs and are all part of the {\it Local Irregulars That Trace Luminosity Extremes; The H\,{\sc i} Nearby Galaxy Survey} ({\it LITTLE THINGS}) survey \citep{hunter.etal.2012} which is a survey of nearby, gas rich irregular/dwarf galaxies. All 3 galaxies have high H\,{\sc i} to stellar disk mass ratios M(HI)/M(star)$>$1 (Table~1) but very low SFRs with values ranging from 0.001 to 0.005 \citep{hunter.etal.2012}.  Although they appear to be irregular in shape, they clearly have disks as indicated by their rotational to vertical velocity ratios $v_{rot}/\sigma_{z}>1$ \citep{johnson.etal.2015}. These classical low luminosity dwarfs are dark matter dominated and their rotation curves indicate that their dark matter halos have flat cores rather than cuspy profiles \citep{oh.etal.2015}.       

\begin{table*}
\centering
\caption{Galaxy Parameters}
\begin{tabular}{lcccccccccccc}
\hline
\colhead{Galaxy} & \colhead{i} &\colhead{P.A} & \colhead{Distance} & \colhead{Spatial} & \colhead{Galaxy}  & \colhead{Magnitude} & \colhead{R$_{25B}$} & \colhead{M(HI)} & \colhead{M(stars)}\\
\colhead{Name} &  \colhead{Degrees} & \colhead{Degrees} & \colhead{Mpc} & \colhead{Scale} & \colhead{Type} & \colhead{Band} & \colhead{ arcseconds} & \colhead{10$^9$~M$_{\odot}$} & \colhead{10$^9$~M$_{\odot}$}\\
\hline
NGC628    &  7   &   20  &  7.3  & 35.4~pc/$^{\prime\prime}$ & SA(s)c, HII & 9.76, B & 360.0 & 5.2 & 15.0\\ 
NGC6946   & 33   &  243  &  5.9  & 28.6~pc/$^{\prime\prime}$ & SAB(rs)cd;Sy2;HII & 9.61, B & 497.9 & 5.8 & 41.0\\   
NGC3184   & 16   &  179  & 11.1  & 53.8~pc/$^{\prime\prime}$ & SAB(rs)cd  & 10.3, B & 255.0 & 3.9 & 23.4\\
NGC4214   & 25.8 &   16  &  3.0  & 14.5~pc/$^{\prime\prime}$ & IAB(s)m    &  9.7, V & 330.0 & 0.5 & 1.1\\
DDO46      & 28   &   84  &  6.1  & 29.6~pc/$^{\prime\prime}$ & Im         & 13.32, V & 66.0 & 0.21 & 0.03\\
DDO63      & 0    &    0  &  3.9  & 18.9~pc/$^{\prime\prime}$ & IAB(s)m    & 13.22, V & 120.0 & 0.17 & 0.04 \\
DDO187     & 39   &   37  &  2.2  & 10.7~pc/$^{\prime\prime}$ & ImIV-V     & 13.9, V & 54.0 & 0.016 & 0.0083\\

\hline
\end{tabular}
\begin{flushleft}
1.~The inclination angles, position angles and distance parameters have been listed from the {\it THINGS} survey \citep{walter.etal.2008} and the {\it LITTLE THINGS} survey \citep{hunter.etal.2012}. 2.~The galaxy types, magnitudes and sizes D$_{25B}$ have been listed from NED. 3.~The H\,{\sc i} and stellar masses were derived from the {\it THINGS} and {\it LITTLE THINGS} surveys using {\it ellint} as decribed in the text. Similarly, the stellar masses were derived using {\it SPITZER} 3.6$\rm{\mu}$m images.  
\end{flushleft}   
\end{table*}

\subsection{Archival Data}

The H\,{\sc i} maps of NGC 628, NGC 6946, NGC 3184 and NGC 4214 were obtained from {\it THINGS} and have a spectral resolution of either 2.6 $\rm{km~s^{-1}}$ or 1.3 $\rm{km~s^{-1}}$ and spatial resolution of $\sim6\arcsec$ in the robust weighting scheme. This is important for obtaining a good H\,{\sc i} surface density profile. For example in NGC628 (Table~1) the H\,{\sc i} images have a spatial resolution of $\sim6^{\prime\prime}$ or 212~pc assuming a distance of 7.3~Mpc. The H\,{\sc i} velocity dispersions were derived by fitting a single Gaussian function to the H\,{\sc i} emission profiles. For the dwarfs (DDO46, DDO63, DDO187), the H\,{\sc i} data was obtained from {\it LITTLE THINGS} \citep{hunter.etal.2012} which has a spectral resolution of either 2.6~$\rm{km~s^{-1}}$ or 1.3~$\rm{km~s^{-1}}$ and a spatial resolution of $\sim6\arcsec$, which matches the resolution of {\it THINGS}. To derive the stellar masses we used the mid-infrared (MIR, 3.6$\rm{\mu}$m) {\it Spitzer} Infrared Array Camera (IRAC) images of the galaxies from the Spitzer Infrared Nearby Galaxies Survey (SINGS) \citep{kennicutt.etal.2003}.  

We have not included the molecular hydrogen gas (H$_2$) masses for the galaxies in this study mainly because we are interested in the dynamical masses of the outer disks of galaxies, where there is very little H$_2$ gas. These regions lie beyond the R$_{25}$ radius, which is the radius at the optical 25 mag arcsec$^{-2}$ isophote. The molecular gas in the galaxies NGC628, NGC6946, NGC3184 and NGC4214 lie well within their R$_{25}$ radii \citep{leroy.etal.2009}. Molecular gas is not detected in DDO63 \citep{leroy.etal.2009}, DDO46 or DDO187 \citep{taylor.etal.1998}.

\subsection{The H\,{\sc i} and stellar surface densities}  

To determine the disk surface density for the H\,{\sc i} disk we used the miriad task {\it ellint} to determine the mean and total flux within annuli of width equal to the mean beamwidth of the observations \citep{sault.etal.1995}. The annuli were centered at the optical center of the galaxy and we used a fixed inclination angle and position angle (Table~1), which are the same as that of {\it THINGS} \citep{walter.etal.2008}. The H\,{\sc i} mass was calculated using the relation M(HI)=$1.4\times 2.36\times10^{5}{D_{Mpc}}^{2}S_{v}$ where $S_v$ is the flux in units of Jy~km~s$^{-1}$ and the multiplicative factor of 1.4 is the correction for including helium (Table~1). 

To determine the stellar mass we used a mass to light (M/L) ratio 0.5 \citep{schombert.mcgaugh.2014} (Table~1). Total luminosities were determined directly from archival IRAC images using
the techniques outlined in \citet{schombert.mcgaugh.2014} from the 3.6$\rm{\mu}$m curve of growth. Note that the H\,{\sc i} mass is an order of magnitude larger than the stellar mass for the dwarf galaxies but is comparable to the stellar mass in the large spiral galaxies (Table~1). The stellar 3.6$\rm{\mu}$m surface brightness at the outermost disk radii lies between 24 to 26 mag/arcsec$^2$ for the dwarf galaxies but for the larger disk galaxies it varies between 21 to 27 mag/arcsec$^2$. However, in all the galaxies the surface brightness profile has a decreasing trend in the outer disk (Figures 2 and 3) and the stellar disk mass surface density appears to be falling rapidly.  

\begin{figure*}
\centering{
\includegraphics[scale=0.30]{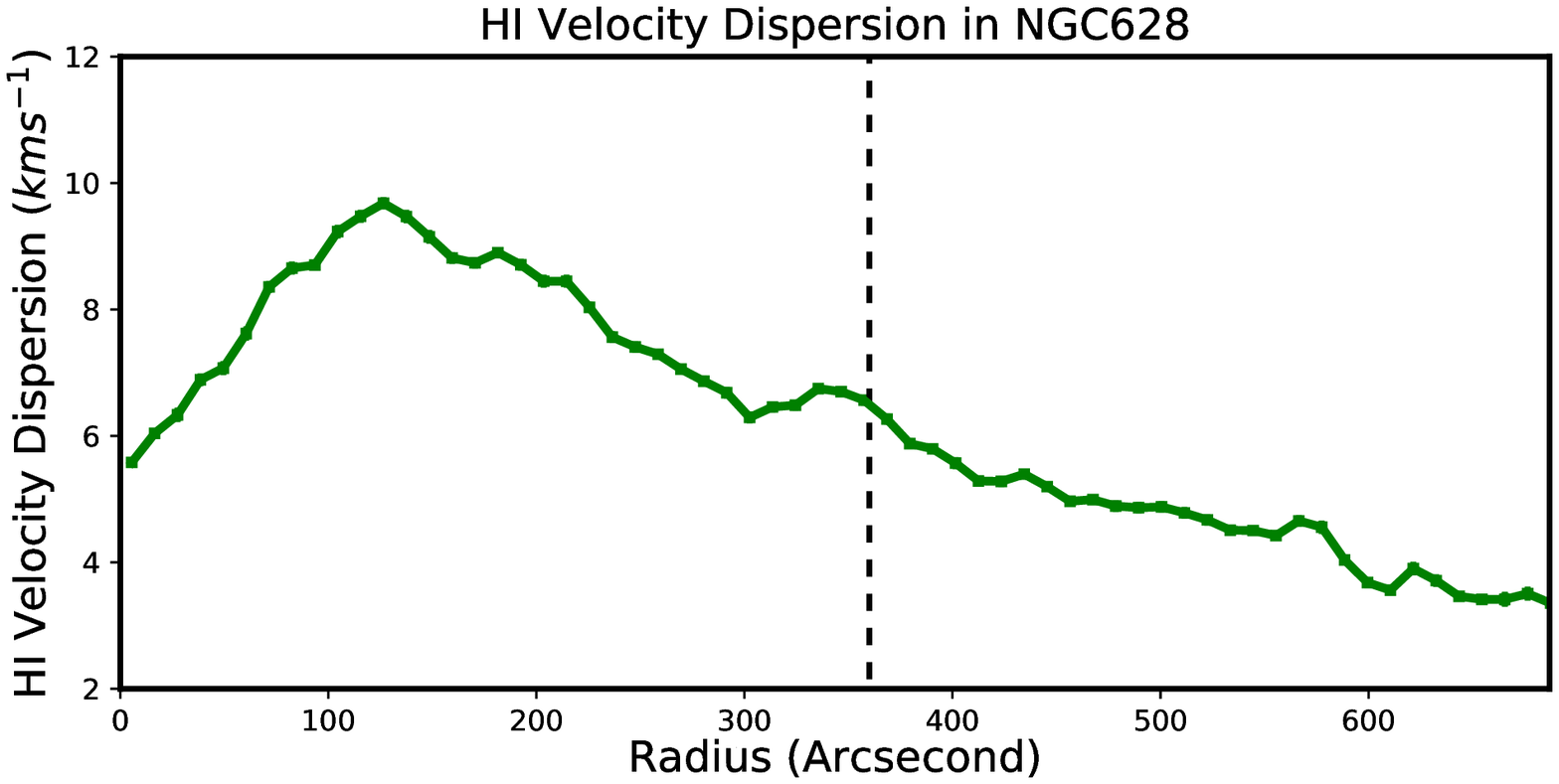}
\hspace{1cm}
\includegraphics[scale=0.30]{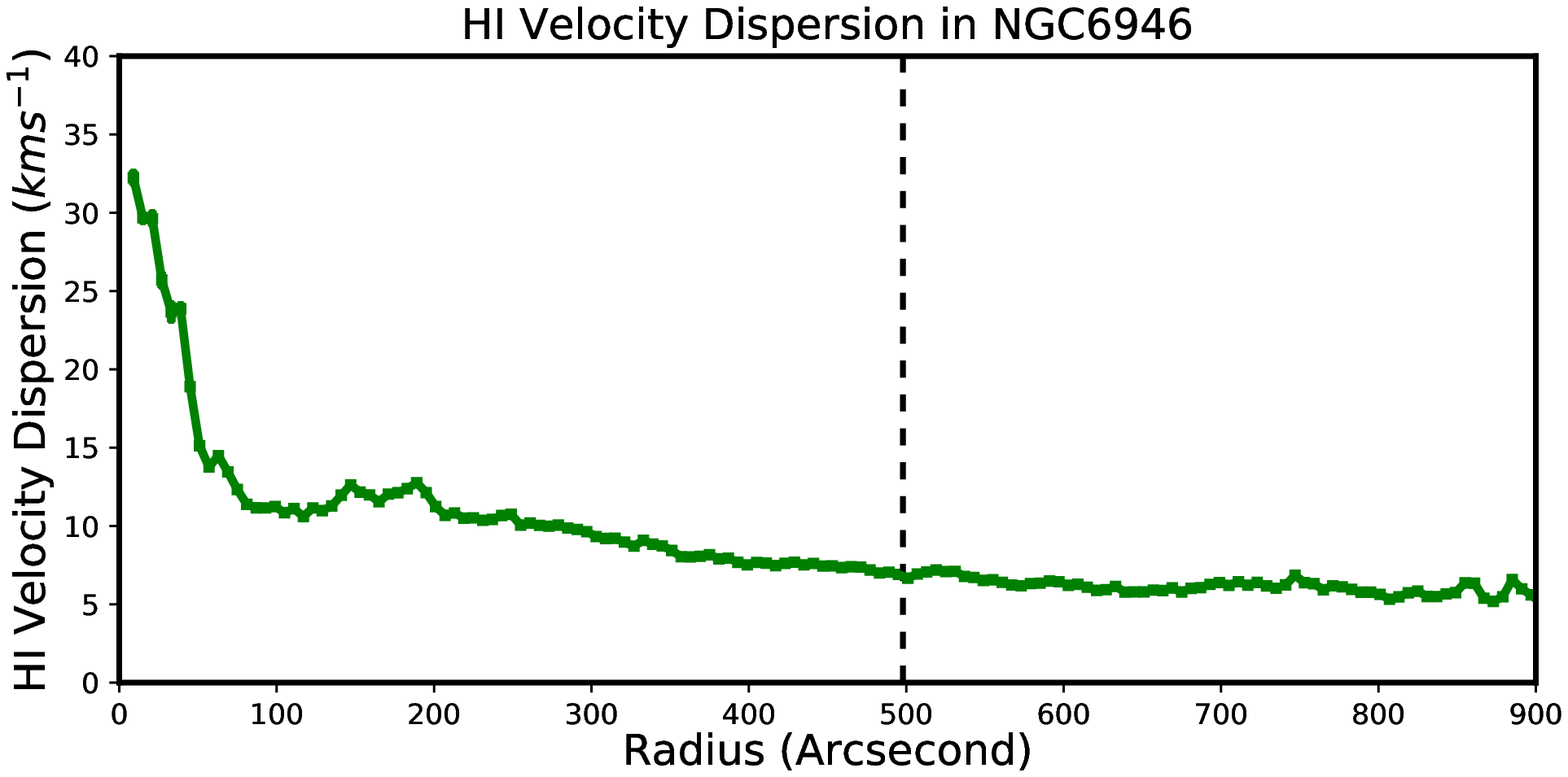}
\includegraphics[scale=0.30]{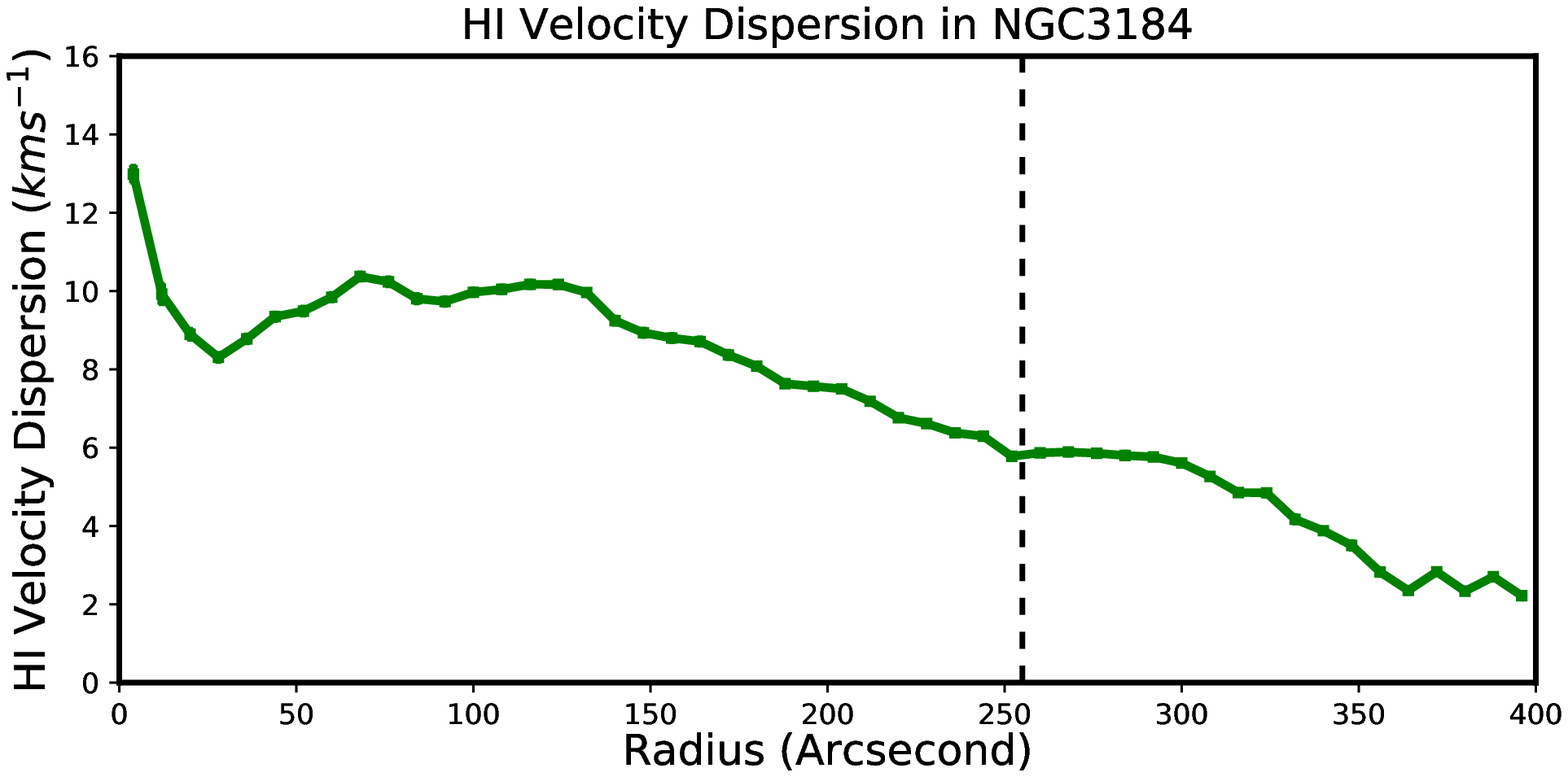}
\hspace{1cm}
\includegraphics[scale=0.30]{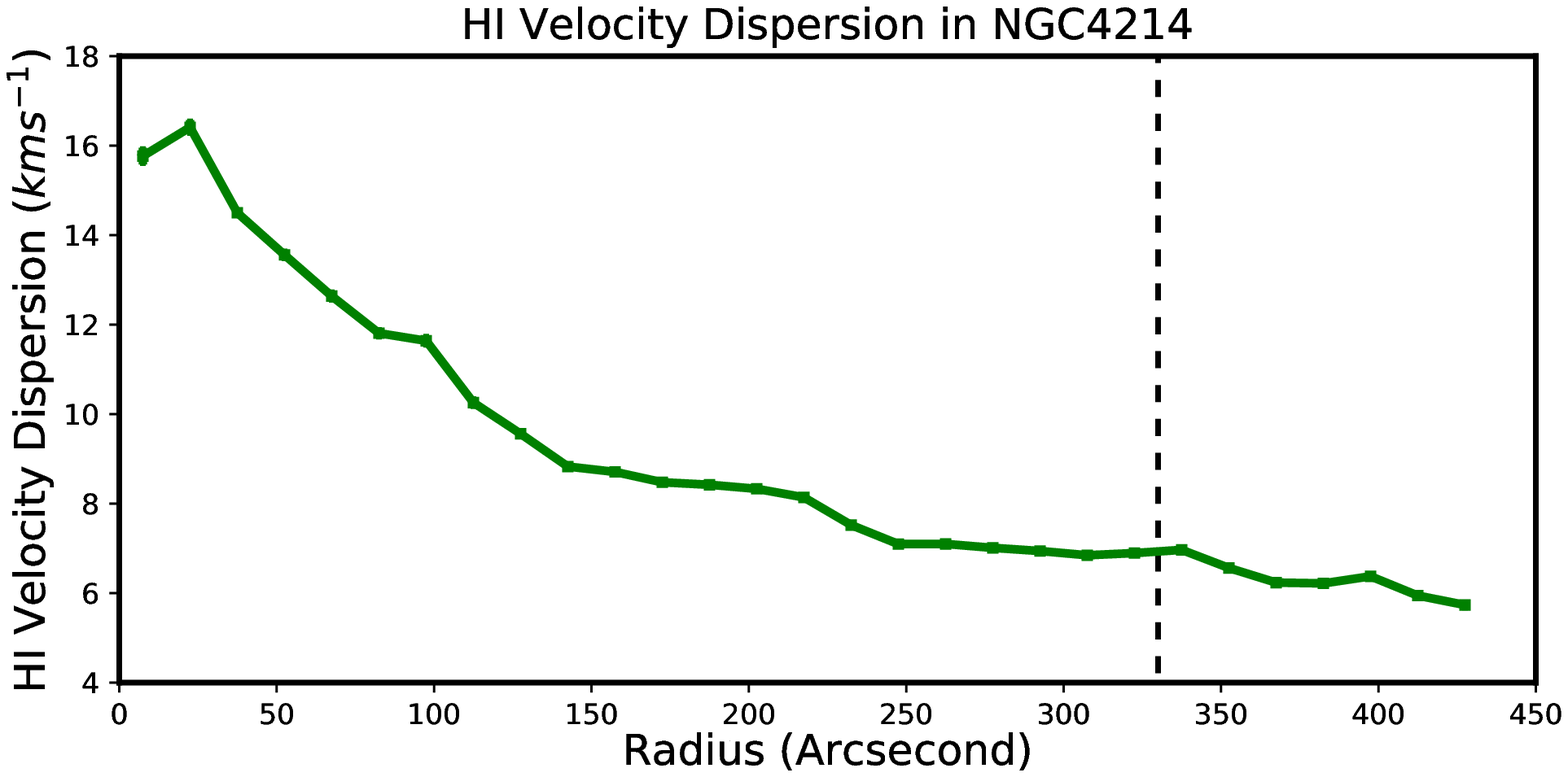}
\includegraphics[scale=0.30]{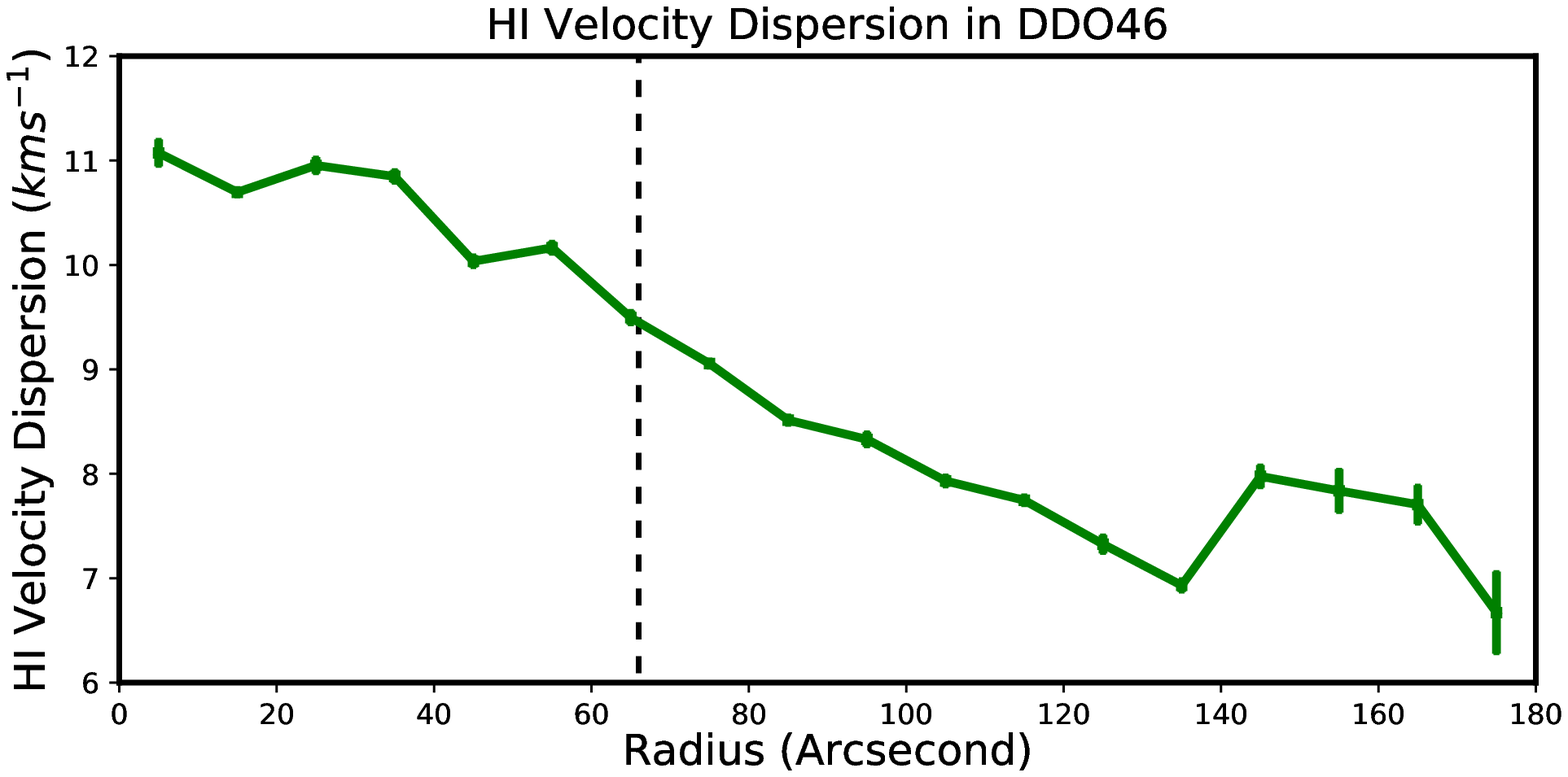}
\hspace{1cm}
\includegraphics[scale=0.30]{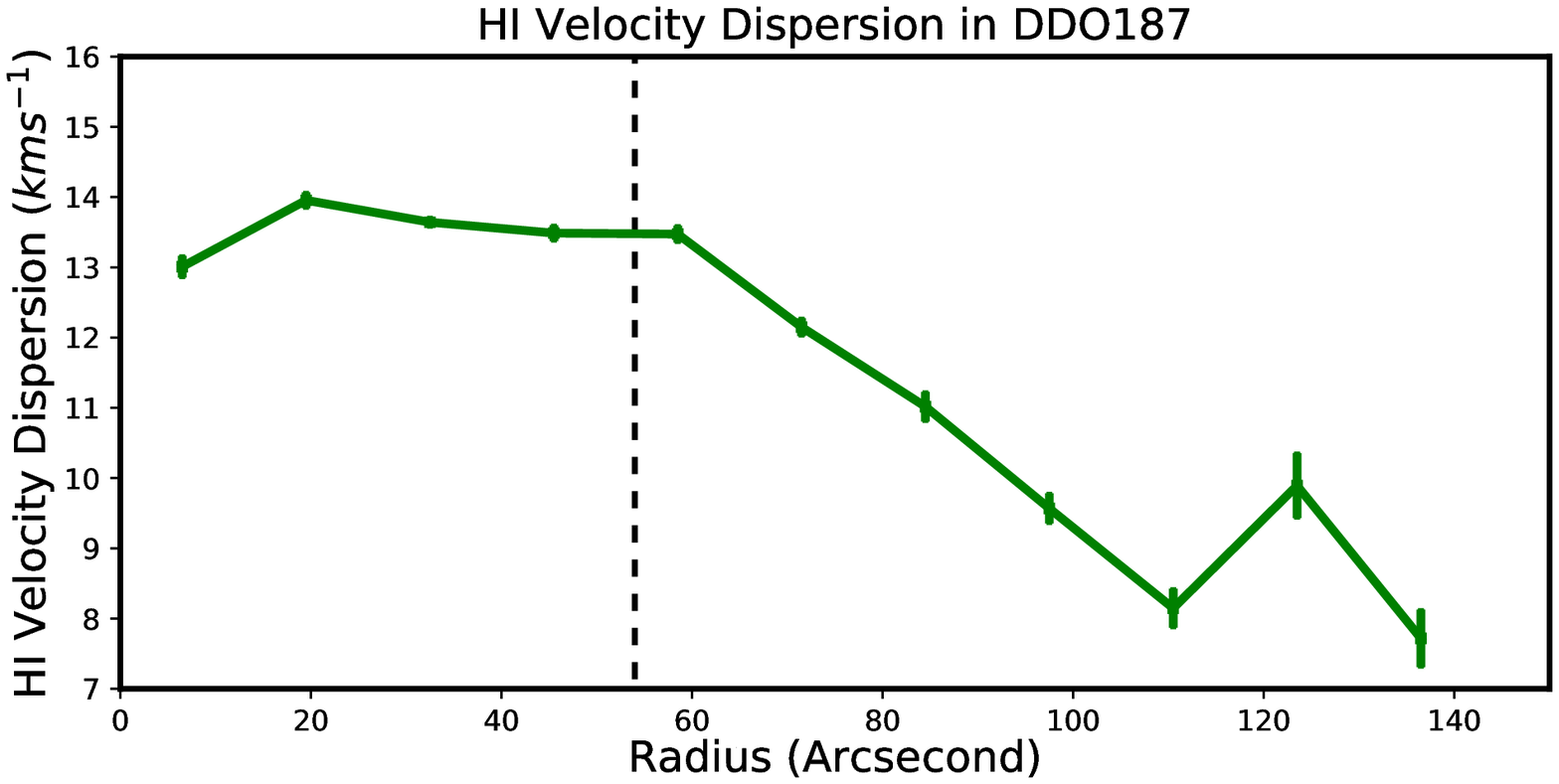}
\includegraphics[scale=0.30]{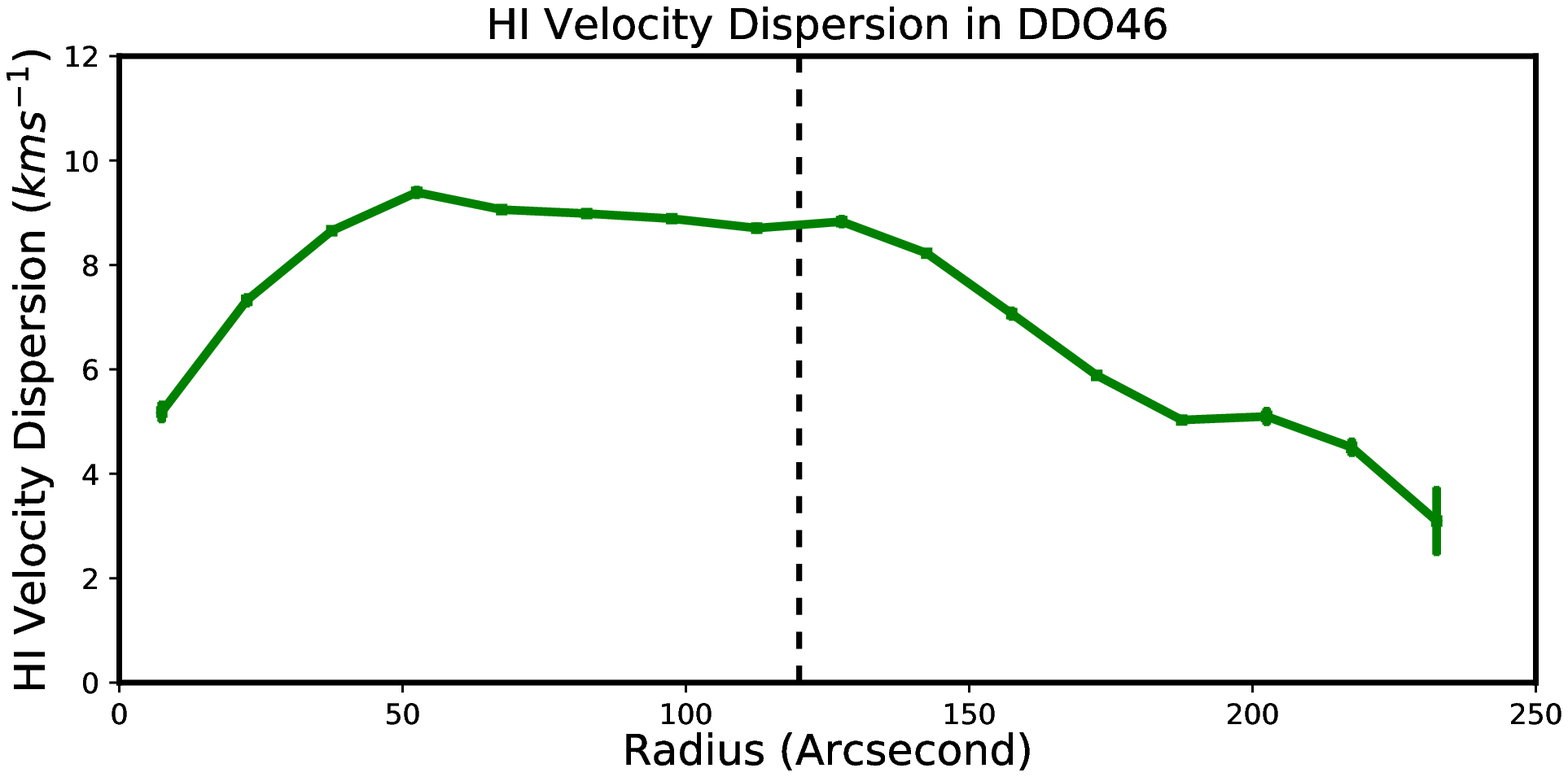}
\caption{(a)~Shown above is the radial variation of the H\,{\sc i} velocity dispersion across the galaxy disks. The vertical dashed line marks the $R_{25}$ radius for the galaxy (see Table~1).}
}
\end{figure*}

The figures in section~4 show the surface mass density of H\,{\sc i} and stars ($\Sigma(HI)$, $\Sigma(s)$) for our sample. The stellar disk lies well within the $R_{25}$ radii for all the galaxies except for DDO187. The H\,{\sc i} gas disk is the most extended component, reaching out to twice the $R_{25}$ radius for nearly all the galaxies except NGC4214.    

\subsection{The H\,{\sc i} velocity dispersion across the disks}

We followed the procedure outlined in \citep{ianja.etal.2012} and \citep{ianjamasimanana.etal.2017} while deriving the radial variations of the H\,{\sc i} velocity dispersion accross the H\,{\sc i} disk of the galaxies. In summary, we line-up individual velocity profiles to the same reference velocity and co-add them to get azimuthally-averaged high signal-to-noise (S/N) profiles. Note that only individual profiles with a S/N better than 3 times the rms noise level in a single channel map were coadded. We then fit the high S/N stacked profiles with a single Gaussian function. The width of the fitted Gaussians represent the velocity dispersion of the bulk of the H\,{\sc i} gas.  As studied in detail in \citet{ianjamasimanana.etal.2017} (see also \citet{ianja.etal.2012} and \citet{mogotsi.etal.2016}), the stacking  procedure outlined above effectively reduce the effects of noise in the derivation of the velocity dispersion. In addition, a Gaussian fit to the profile gives a more robust estimate of the velocity dispersion as opposed to the more straightforward moment map analysis. This is because moment maps values are more prone to the effects of low-level uncleaned fluxes than single Gaussian fitting values unless a very careful clipping is applied. For a full discussion of this, we refer the reader to \citet{ianjamasimanana.etal.2017}. Note that in their papers \citet{ianja.etal.2012} and  
\citet{ianjamasimanana.etal.2015} separate the HI profiles in terms of broad and narrow Gaussian components. Here, our choice of a single Gaussian fit is motivated by the fact that we are mainly interested in the outer disks of galaxies where the contributions of the narrow components to the global shapes of the azimuthally averaged velocity profiles are not important  (for details see \citet{ianjamasimanana.etal.2015}). Thus a separation into narrow and a broad Gaussian components are not relevant for the purpose of this work. Figure~2 shows the radial variation of the velocity dispersion across the galaxy disks. Note that the large galaxies, and especially NGC6946, clearly show non-declining H\,{\sc i} velocity dispersion in the outer parts of their disks.           

The nearly face-on orientation of the galaxies in our sample means that the H\,{\sc i} velocity dispersion closely traces the velocity in the z direction ($\sigma_{\rm{H\,{\textsc i}}}$). We use equation~[13] with the respective inclination angles (Table~1) to obtain $\sigma_{\rm{zH\,{\textsc i}}}$. As reported in the literature, the velocity dispersions have a much shallower profile than the SFR density profiles even in the outer disks where star formation is absent or unimportant. This indicates that mechanisms other than star formation are also important to drive the linewidth of the H\,{\sc i}.

\begin{figure*}
\centering{
\includegraphics[scale=0.40]{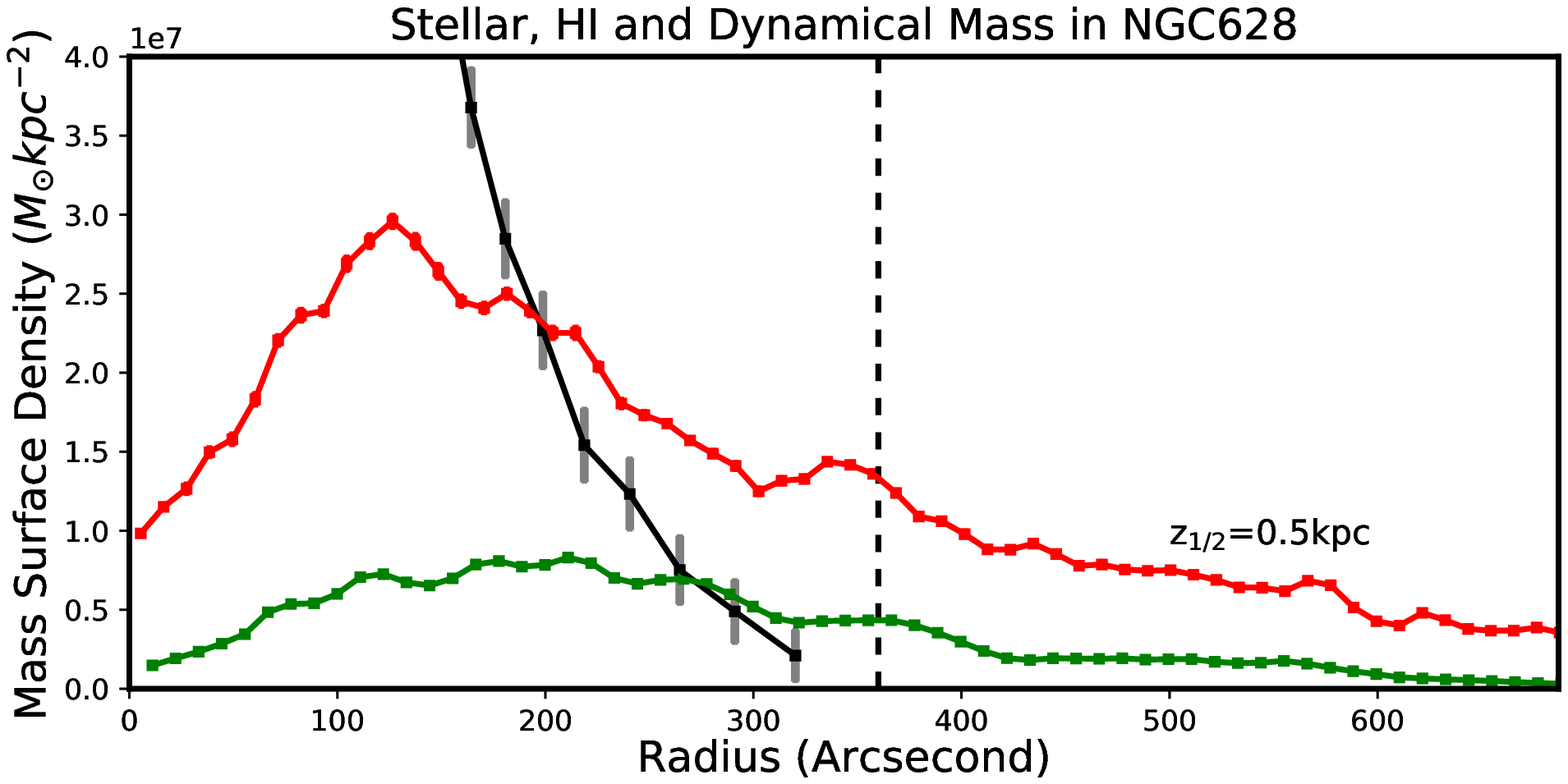}
\hspace{1cm}
\includegraphics[scale=0.40]{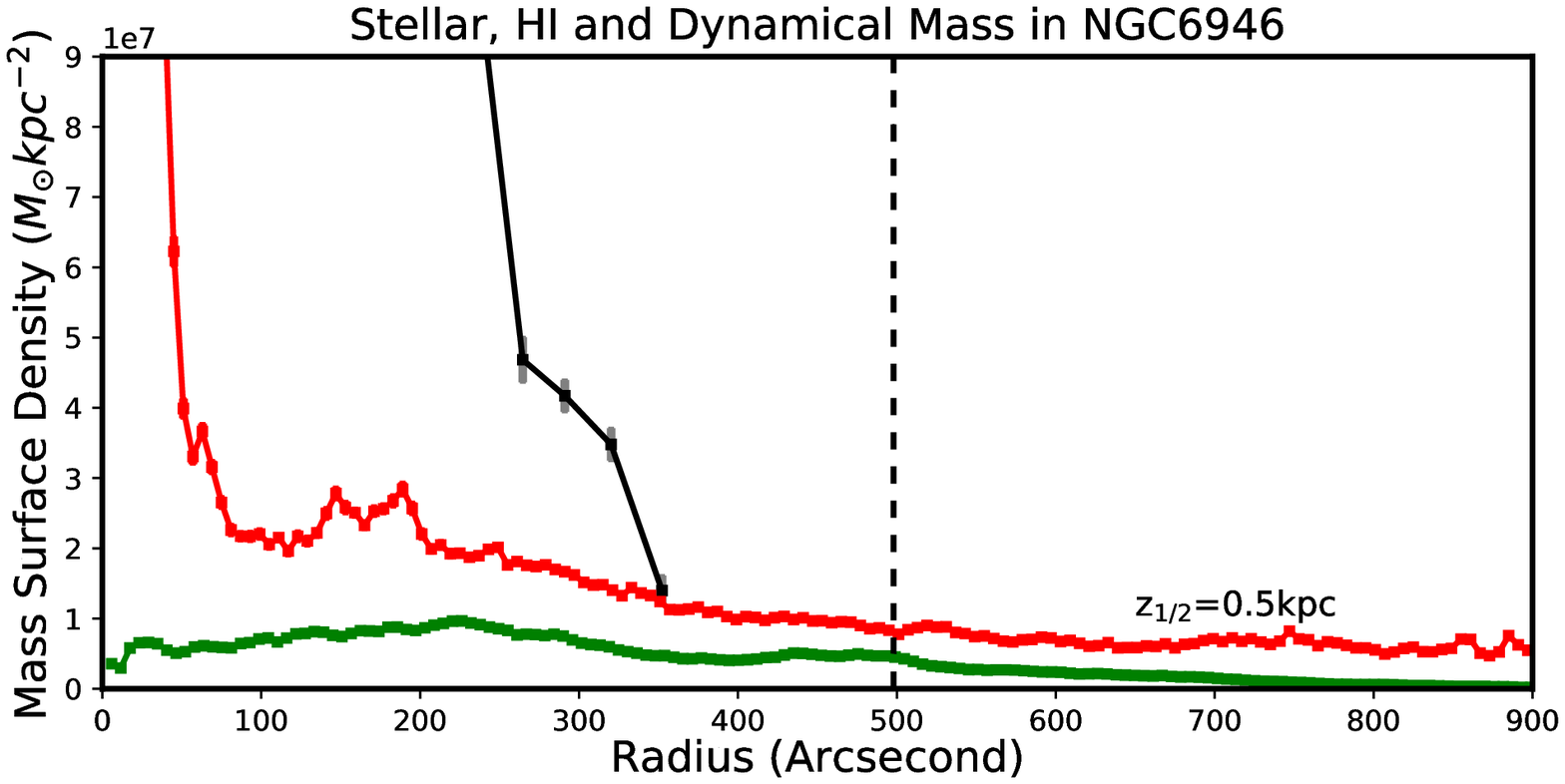}
\includegraphics[scale=0.40]{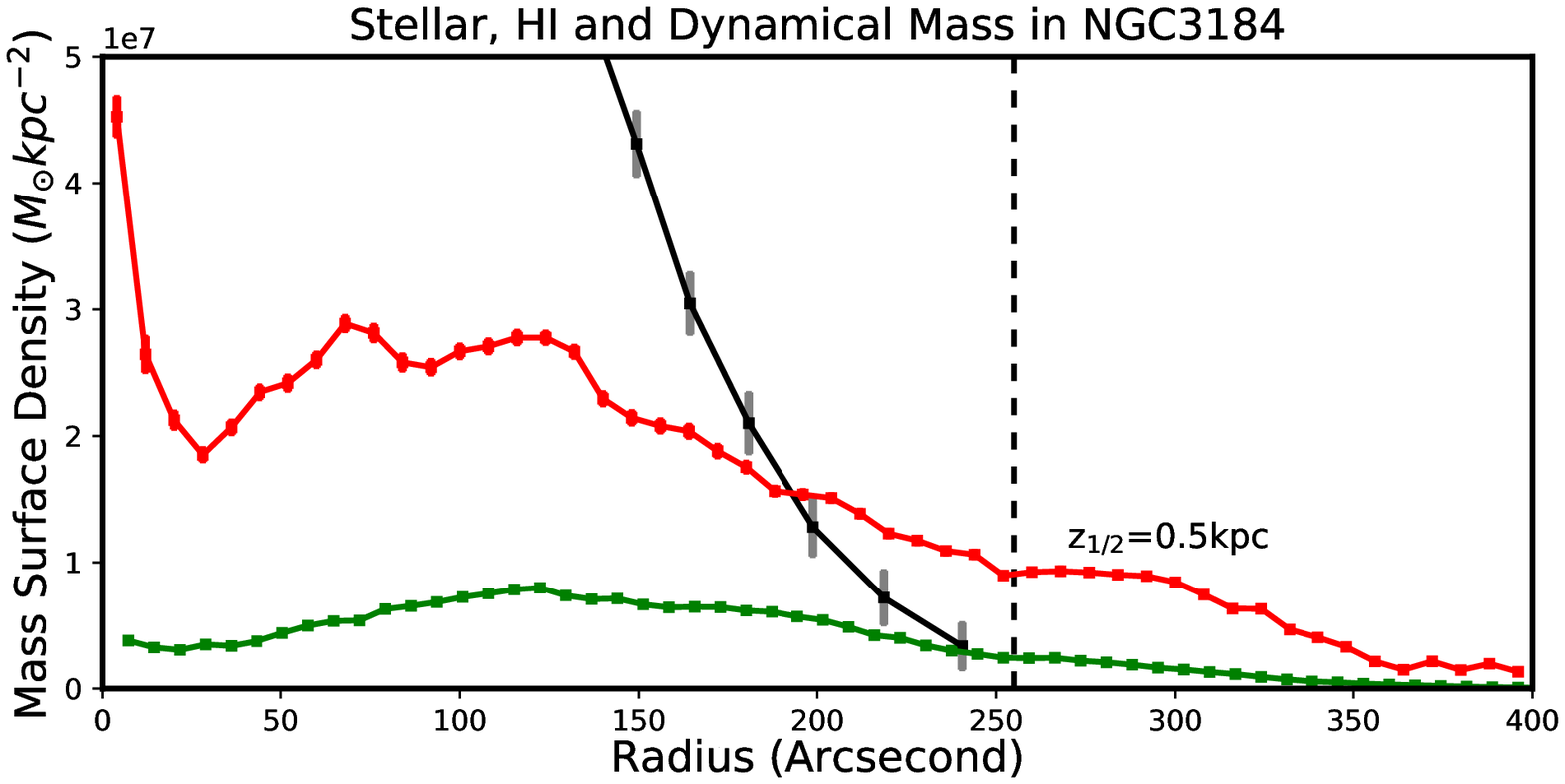}
\hspace{1cm}
\includegraphics[scale=0.40]{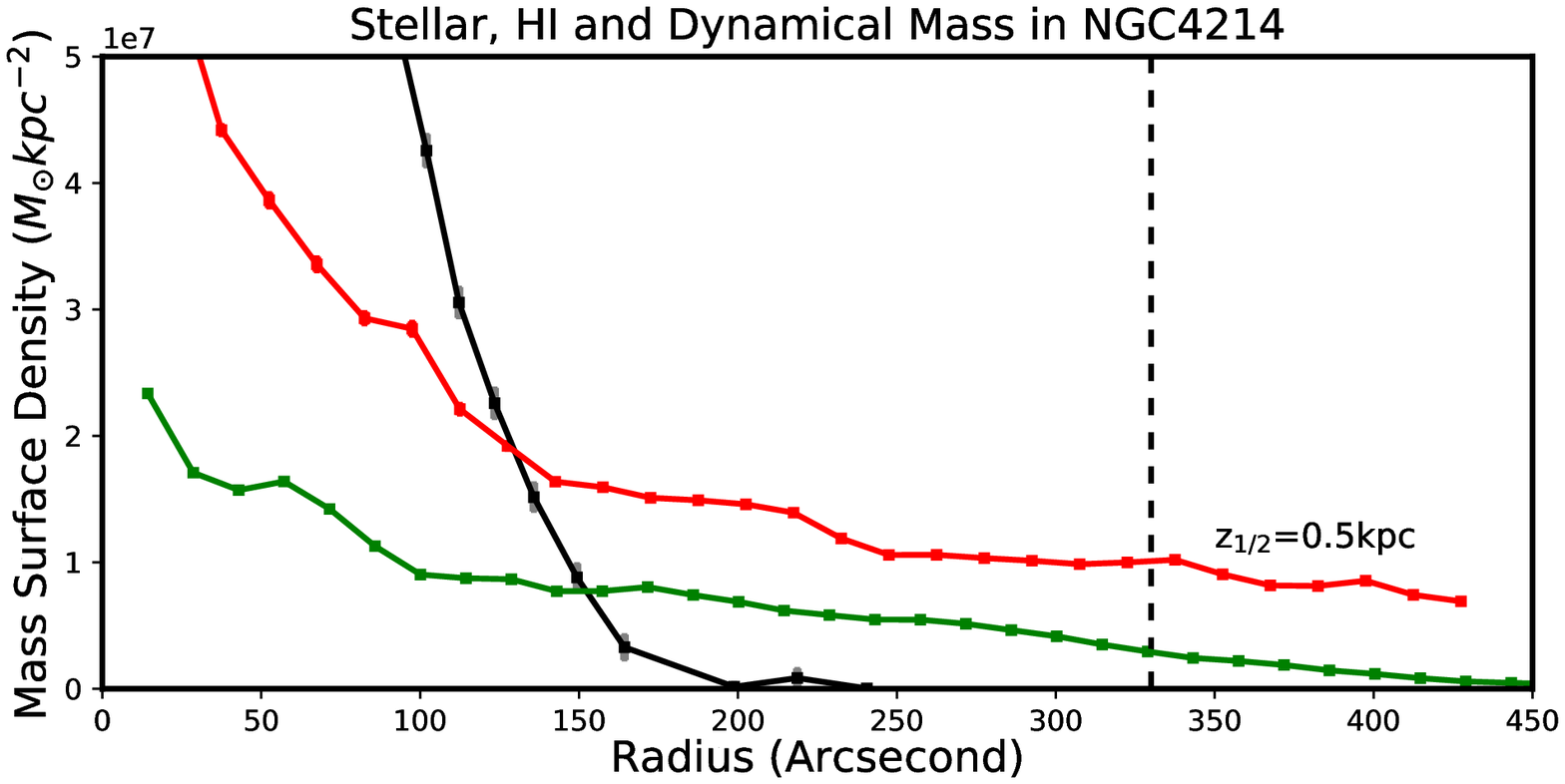}
\caption{(a)~Shown above is the surface mass density of stars $\Sigma(stars)$ (black), H\,{\sc i} $\Sigma(HI)$ (green) and the dynamical mass derived from the H\,{\sc i} velocity dispersion $\Sigma(dyn)$ (red) for the large disk galaxies NGC628, NGC6946, NGC3184, NGC4214. The disk half width $z_{1/2}$ is assumed to be 0.5kpc for all galaxies and the dashed vertical line marks the $R_{25}$ radius of the galaxy.  Note that the dynamical mass density $\Sigma(dyn)$ is comparable to $\Sigma(HI)$ only in regions where there is very little stellar mass, which is usually in the extreme outer radii. All 4 galaxies have M(HI)$<$M(stars).}
}
\end{figure*}

\begin{figure*}
\centering{
\includegraphics[scale=0.40]{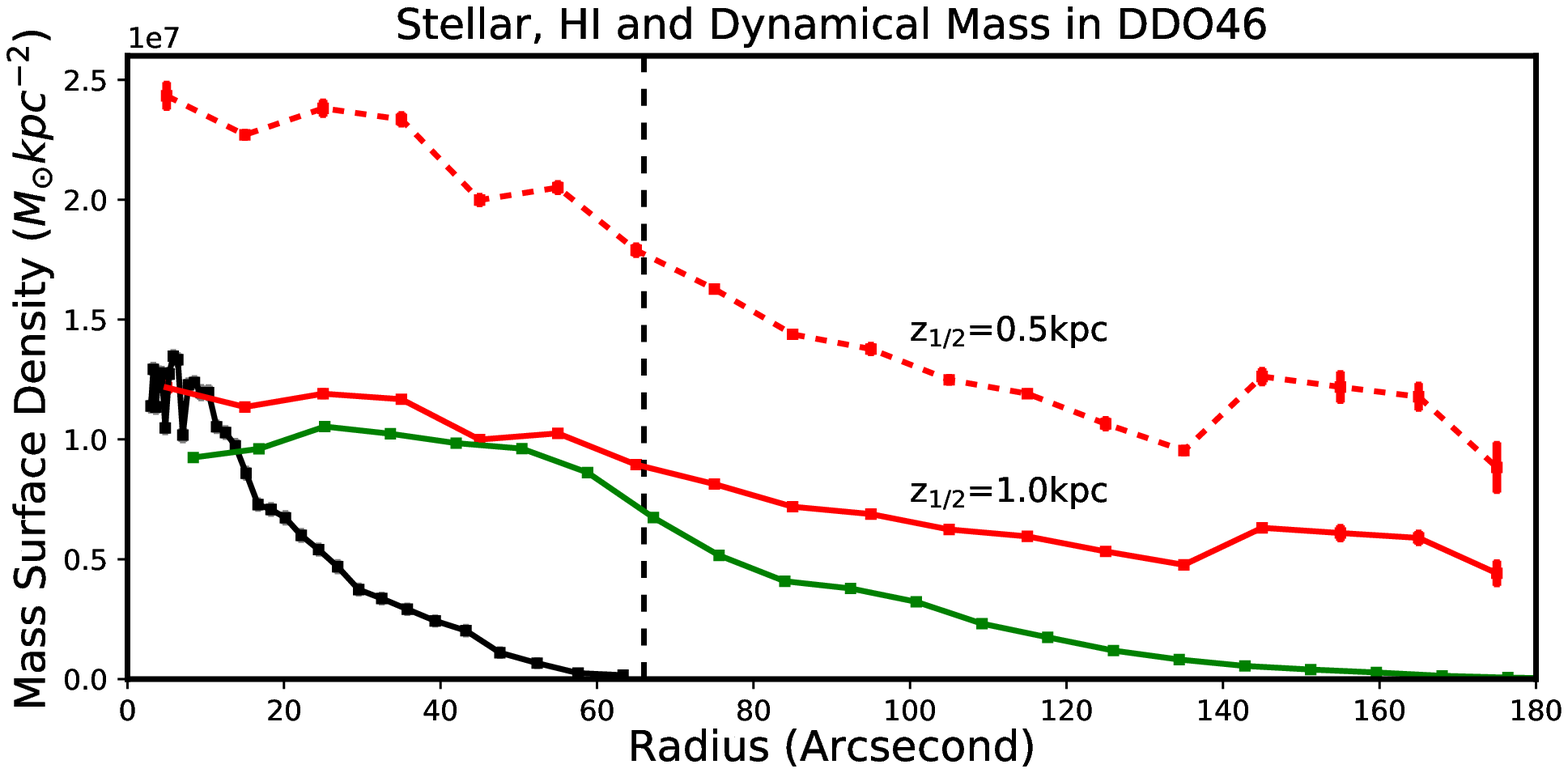}
\hspace{1cm}
\includegraphics[scale=0.40]{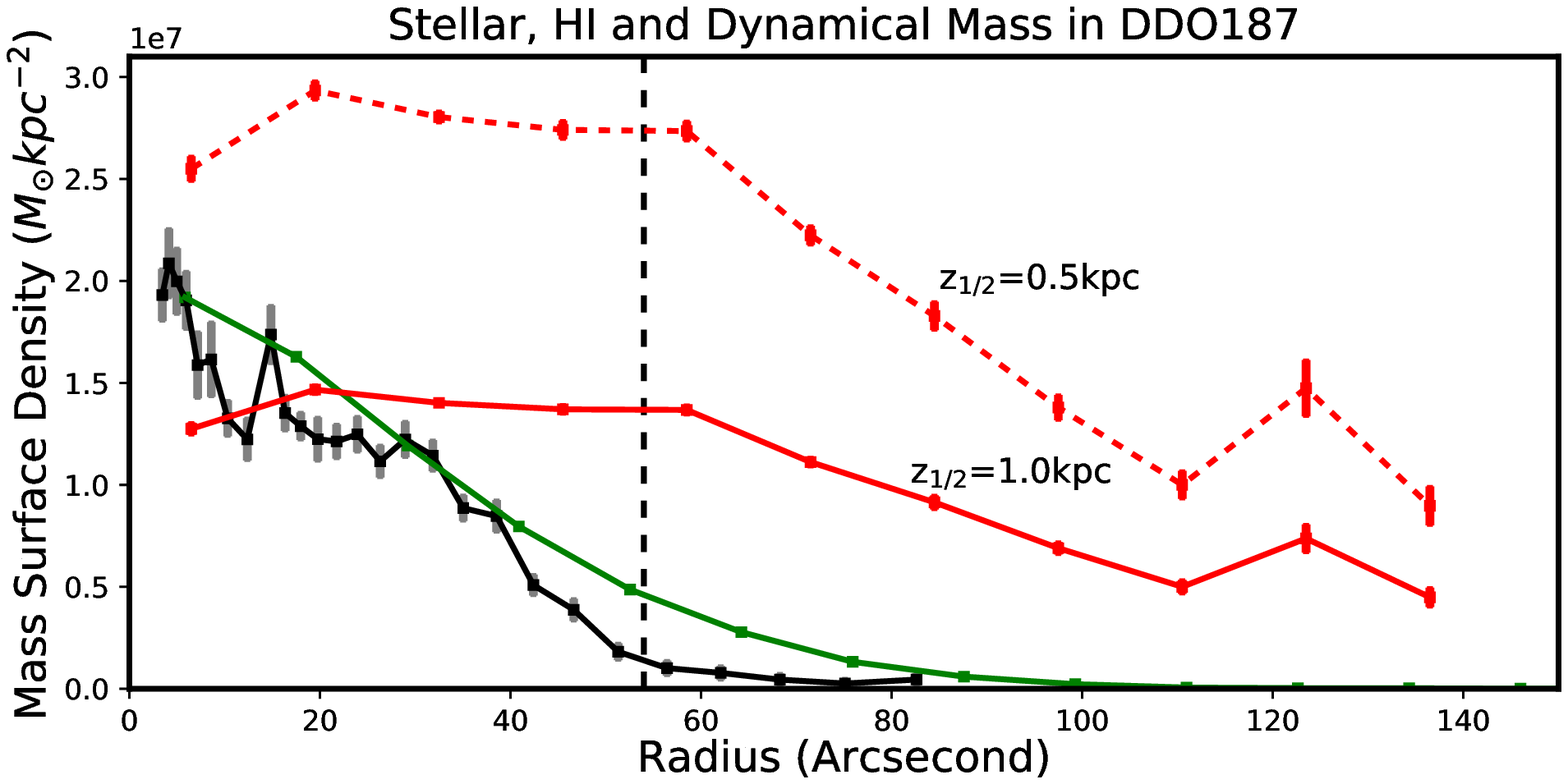}
\includegraphics[scale=0.40]{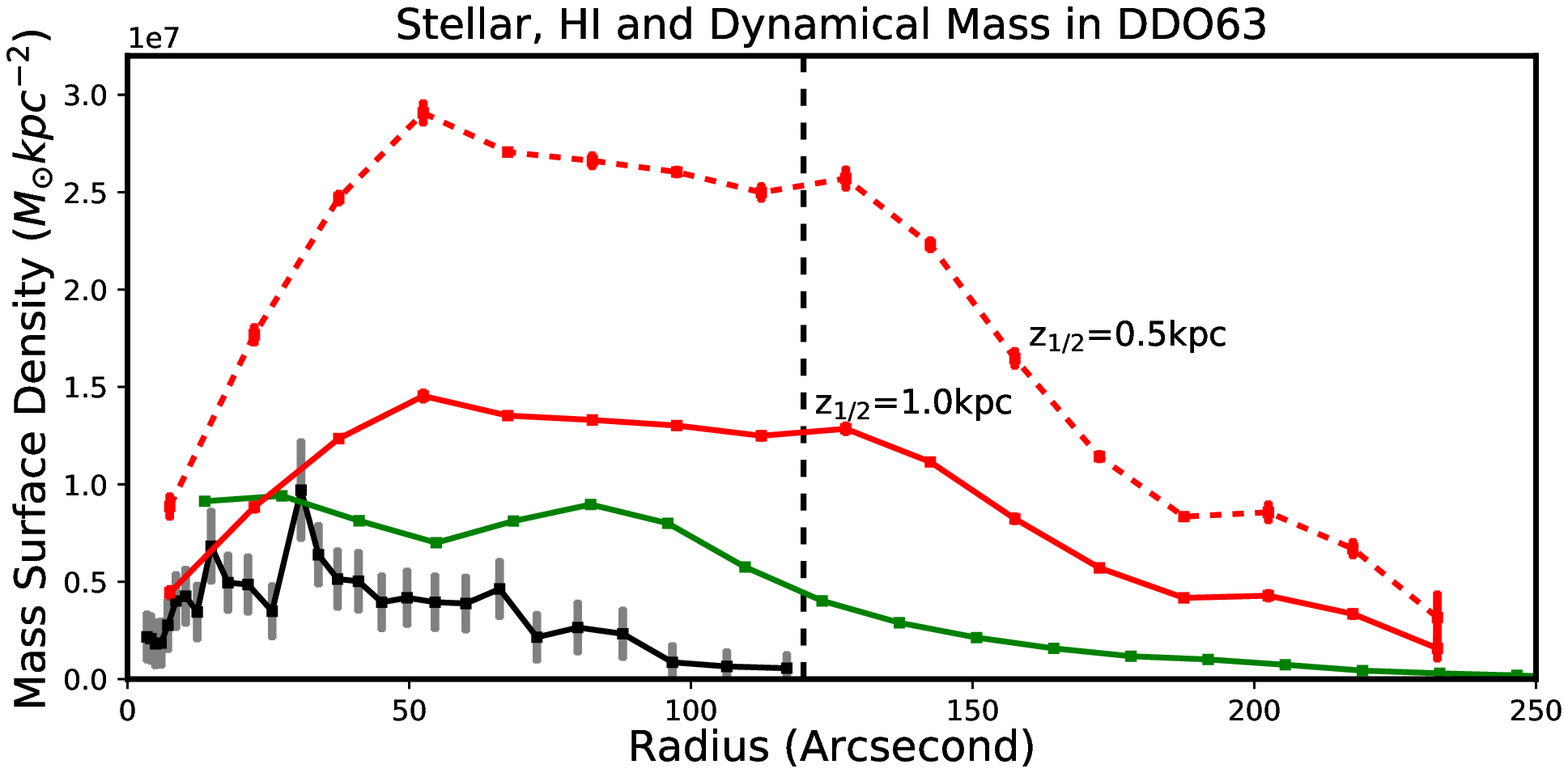}
\caption{(a)~Shown above is the surface mass density of stars $\Sigma(*)$ (black), H\,{\sc i} $\Sigma(HI)$ (green) and the dynamical mass derived from the H\,{\sc i} velocity dispersion $\Sigma(dyn)$ (red) for the gas rich dwarf galaxies DDO46, DDO187 and DDO63. The disk half width $z_{1/2}$ is assumed to be 0.5~kpc and 1~kpc for all galaxies; the dashed vertical line marks the $R_{25}$ radius. Note that the dynamical mass density $\Sigma(dyn)$ is greater than $\Sigma(HI)$ nearly over the entire galaxy disks. }
}
\end{figure*}

{\bf
\begin{table*}
\centering
\caption{Approximate baryonic and dynamical masses in the region $R>R_{25B}$}
\begin{tabular}{c|c|c|c|c|c|c|c}
\hline
Galaxy   & M(H\,{\sc i})               & M(stars)      & $z_{1/2}(HI)$ & M(dyn)              & Relevant  & $\frac{M_{dyn}}{M(HI)+M(stars)}$ & $M_{dyn}$-[M(HI)+M(stars)]\\
Name     & ($M_{\odot}$)       & ($M_{\odot}$) & (kpc)         & ($M_{\odot}$)       & radii     &     &  ($M_{\odot}$)  \\
\hline
NGC628  & $2.05\times10^{9}$  & ....           & 0.5          & $9.11\times10^{9}$  & (1-1.9)$R_{25B}$, (12.7-24.2)kpc & 4.4 & $7.1\times10^{9}$\\
NGC6946 & $2.05\times10^{9}$  & ....           & 0.5          & $9.62\times10^{9}$  & (1-1.8)$R_{25B}$, (14.2-25.6)kpc & 4.7 & $7.6\times10^{9}$\\
NGC3184 & $8.79\times10^{8}$  & ....           & 0.5          & $4.56\times10^{9}$  & (1-1.6)$R_{25B}$, (13.7-21.9)kpc  & 5.2 & $3.7\times10^{9}$\\
NGC4214 & $9.16\times10^{7}$  & ....           & 0.5          & $4.33\times10^{8}$  & (1-1.3)$R_{25B}$, (4.8-6.2)kpc & 4.7 & $3.4\times10^{8}$\\
\hline
DDO46    & $1.23\times10^{8}$  & ....           & 0.5         & $5.71\times10^{8}$  & (1-2.7)$R_{25B}$, (2.0-5.3)kpc &  4.6 & $4.5\times10^{8}$\\
         &                     & ....           & 1.0         & $2.85\times10^{8}$  &                       &  2.3 & $1.6\times10^{8}$\\
DDO63    & $6.44\times10^{7}$  & ....           & 0.5         & $5.30\times10^{8}$  & (1-1.9)$R_{25B}$, (2.3-4.3)kpc & 8.2 & $4.7\times10^{8}$\\
         &                     & ....           & 1.0         & $2.65\times10^{8}$  &                       & 4.1   & $2.0\times10^{8}$\\
DDO187   & $5.20\times10^{6}$  & $0.96\times10^{6}$ & 0.5     & $8.81\times10^{7}$  & (1-2.6)$R_{25B}$, (0.58-1.5)kpc & 14.3 & $8.2\times10^{7}$\\
         &                     &                & 1.0         & $4.41\times10^{7}$  &                       & 7.2 & $3.8\times10^{7}$\\
\hline
\end{tabular}
\tablecomments{The H\,{\sc i}, stellar and dynamical masses are derived for the regions $R>R_{25B}$. The lower and upper radii are listed in column 6, and the $R_{25B}$ radii are given in Table~1.}
\end{table*}
} 

\section{Results}

\subsection{Dynamical and H\,{\sc i} Masses of the Outer Disks}
One of the main aims of this paper is to investigate what gravitationally binds the extended H\,{\sc i} disks of galaxies in regions where there is no visible stellar disk. For example Figures~3 and 4 show that the stellar disks traced by 3.6$\rm{\mu}$m emission, have smaller disk radii than the H\,{\sc i} disks. Also, the stellar disks lie within the $R_{25}$ radii for all the galaxies except for DDO187. Within the $R_{25}$ radii of the large galaxies (Figure~1), the stellar mass surface density $\Sigma_s$ is several magnitudes higher than $\Sigma_{\rm{H\,{\textsc i}}}$. The gravity of the stellar disk maintains the vertical structure of the H\,{\sc i} disk. The star formation within this region can give rise to turbulence in the gas which can contribute to the gas pressure which helps to balance the gravitational force of the disk. 

But in the regions $R>R_{25}$ where the stellar disk surface density $\Sigma_{s}$ is very low, there must be some other mass to gravitationally bind the H\,{\sc i} disk since just the self gravity of the H\,{\sc i} cannot maintain the vertical structure. We have applied our analytical results discussed in Section~2 to these regions i.e. radii $R>R_{25}$, by assuming vertical equilibrium and applying equation~[11], i.e. $\Sigma_{dyn}=\frac{{\sigma_{zHI}}^{2}}{\pi Gz_{g0}}$. In Figure~3, we have compared $\Sigma_{dyn}$ with $\Sigma_{\rm{H\,{\textsc i}}}$ for the normal galaxies using the vertical disk scale height $z_{1/2}=0.5$~kpc. Figure~4 shows the same for the dwarf galaxies but for two vertical disk scale heights, $z_{1/2}=0.5$~kpc and $1$~kpc. Observations of edge-on galaxies suggests that $z_{1/2}=0.5$~kpc is a good approximation for the vertical height of the H\,{\sc i} layer in large disk galaxies \citep{zschaechner.rand.2015}. But dwarf galaxies have puffier H\,{\sc i} disks, hence $z_{1/2}=0.5$~kpc and $1$~kpc are suitable values for this pilot study \citep{kamphuis.etal.2011}. The assumption here is that the dynamical mass has a scale height similar to the H\,{\sc i} vertical scale height. The $\Sigma_{dyn}$ decreases for larger vertical disk heights, which is to be expected as the smaller $z_{1/2}$ values means larger disk gravity and hence larger dynamical mass (Figure~4). 

From Figure~3 it is clear that $\Sigma_{dyn}$ as traced by the H\,{\sc i} velocity dispersion is at least 1.5 to 2 times the value of $\Sigma_{\rm{H\,{\textsc i}}}$ in the region $R>R_{25}$ for the galaxies NGC628, NGC3184 and NGC4214. For NGC6946, the difference between $\Sigma_{dyn}$ and $\Sigma_{\rm{H\,{\textsc i}}}$ is not very significant but since the disk is very extended, the total dynamical mass beyond $R_{25}$ does become important. Table~2 shows the H\,{\sc i}, stellar and dynamical masses determined for the regions $R>R_{25}$. The upper radii for mass estimates are listed in column 6, and the $R_{25}$  radii (in the B band) are given in Table~1. The upper radius is limited by the signal to noise of the H\,{\sc i} velocity dispersion which traces the dynamical mass. In the case of the gas rich dwarf galaxies, M(dyn) is much larger than M(\rm{H\,{\textsc i}}) for nearly all radii (Figure~4) and especially for $R>R_{25}$.  For radii $R<R_{25}$, the 
M(\rm{H\,{\textsc i}}) is comparable to or larger than the stellar mass M(stars). This is not surprising as it is well known that the gas to stellar mass ratios are much higher for dwarfs i.e. M(HI)/M(stars)$>>$1 (Table~1). But for the outer radii ($R>R_{25}$), depending on the assumed vertical scale heights of the disks, the M(dyn) can vary from 2 to 14 times the value of M(\rm{H\,{\textsc i}}). The most striking case is DDO187, where M(dyn) is several times the value of M(\rm{H\,{\textsc i}}) in the region $R>R_{25}$ for both $z_{1/2}$=0.5 and 1~kpc (Table~2). 

\subsection{Comparison of the dynamical to baryonic disk masses for radii $R>R_{25}$}

To compare the dynamical and baryonic masses (M(HI)+M(stars)) of our galaxies in the outer disk regions, we have summed over the surface densities  $\Sigma_{dyn}$ and [$\Sigma_{HI}$+$\Sigma_{s}$] in the outer radii (Table~2). For all the galaxies except DDO187, we find that there is no stellar mass visible in the 3.6$\rm{\mu}$m NIR images at radii $R>R_{25}$. Hence, they have M(baryonic)=M(HI) for these regions. For the normal, large galaxies NGC628, NGC6946, NGC3184 and NGC4214, assuming a disk half thickness or height value of $z_{1/2}$=0.5~kpc the M(dyn)/M(baryonic) varies between 4.4 and 5.2. But for the dwarf galaxies, since their stellar disks are more diffuse and maybe thicker, we have calculated M(dyn)/M(baryonic) for $z_{1/2}$=0.5 and 1~kpc (Table~2). It varies between 4.6 to 14.3 for $z_{1/2}$=0.5, but varies between 2.3 to 7.2 for $z_{1/2}$=1~kpc. As the disk thickness is increased, the difference between the dynamical and baryonic masses becomes smaller, which follows from the form of the expression for $\Sigma_{dyn}$ (equation~[11]). The M(dyn)/M(baryonic) is the largest for the gas rich dwarf galaxy DDO187. It is clear from Table~2 that for some of the dwarfs the non-luminous mass associated with the disks is very significant and comparable to or more than the stellar disk mass of the galaxy (Table~1).

An estimate of the non-luminous mass associated with the disks for the region $R>R_{25}$ is shown in the last column of Table~2. The values of $M_{dyn}$-[M(HI)+M(stars)] are of the order of $\sim 10^{9}~M_{\odot}$ for the larger galaxies (NGC628, NGC6946, NGC3184) but around $\sim 10^{7} - 10^{8}~M_{\odot}$ for the smaller/dwarf galaxies. In fact for the dwarf galaxies this mass is comparable to the  stellar disk masses. The smaller galaxies are also the ones that are more gas rich and lower in luminosity. Hence, the outer disks of dwarf and low luminosity galaxies maybe dark matter dominated as discussed in the next section. However, it must be noted that the mass in the disk region $R>R_{25}$ is in no way comparable to the total halo masses of the galaxies, which are typically 10 to 100 times the stellar masses in bright to low luminosity galaxies. Our present sample consists of nearly face on galaxies for which reliable rotation curves cannot be obtained and hence we cannot estimate the dark halo masses.    
  
\subsection{The Non-declining Dynamical Masses of the Outer Disks}
One of the surprising aspects of the H\,{\sc i} velocity dispersion profiles of nearby galaxies is that it does not always fall rapidly to zero in the outer disks \citep{ianjamasimanana.etal.2017,ianjamasimanana.etal.2015} (Figure 2) and hence the dynamical mass also does not decline (Figures 3 and 4). This is especially true in the case of the large galaxies NGC628, NGC6946 and NGC4214 and the dwarfs DDO46 and DDO187, all of which clearly have non-declining disk dynamical mass curves. This is somewhat similar to the rotation curves of galaxies where the non-declining, flat rotation curves at large galactic radii indicates the presence of halo dark matter \citep{rubin.etal.1985,mcgaugh.etal.2001,mcgaugh.2014}. Similarly, for the extended H\,{\sc i} disks, the large differences between the disk dynamical masses and baryonic masses of face-on galaxies may indicate the presence of non-luminous disk mass extending out to radii well beyond the stellar disks. Deeper and more sensitive H\,{\sc i} observations with upcoming telescopes such as the ngVLA or SKA, will help us probe these outer disk regions and put better contraints on the H\,{\sc i} disk sizes of galaxies \citep{blyth.etal.2015}.

\section{Implications}

One of the most important results of this study is that the baryonic mass surface density maybe lower than the dynamical mass surface density in the outer disks of galaxies, which implies that there is not enough baryonic mass (in the form of stars and H\,{\sc i} gas) to bind the disk structure beyond $R_{25}$. This is clear from Table~2 which shows that $\frac{M_{dyn}}{M(HI)+M(stars)}>1$ in the region $R>R_{25}$. This could mean two things, (i)~there is low luminosity stellar mass in the outer disks which provides the gravity to maintain the vertical structure of the extended H\,{\sc i} disks or (ii)~there is dark matter associated with the disks. 

For the first case, most of the low luminosity mass has to be in the form of low mass stars that are not detected in near-infrared (NIR) observations. Deeper NIR observations could reveal more of this disk mass. However, the stellar mass profiles in Figures~3 and 4 indicate that the stellar mass distributions fall sharpy near the $R_{25}$ radius for all the galaxies. So a low luminosity stellar disk providing the gravity for maintaining the disk in the region $R>R_{25}$ is unlikely. 

If dark matter provides the gravitational force for binding the disk, it can have two possible origins. First of all, the dark matter could be the inner, denser part of the galaxy halo. In this scenario, the halo has a flattened or oblate shape and so the inner halo is associated with the disk, which increases the mass surface density of the disk. The corresponding $\Sigma_{dm}$ (equation~[12]) will increase the total mass surface density and support the vertical equilibrium of the H\,{\sc i} disk. Oblate halo shapes have been investigated in earlier studies \citep{olling.1996,becquaert.combes.1997} using the H\,{\sc i} velocity dispersion of nearby edge-on galaxies, but not with face-on galaxies as discussed in this paper. \citet{olling.n4244.1996} applied their method to NGC~4244 and their results suggest that the halo has an axis ratio of $\sim$0.2, which represents a very flat halo and could mean that the dark matter halo mass is associated with the extended H\,{\sc i} disk. The method has also been applied to the Milky Way \citep{olling.merrifield.2000} but the results suggests that the Milky Way halo is not oblate.

Flattened, non-spherical halo shapes may result from higher values of halo spin. Simulations suggest that low luminosity galaxies, such as LSB galaxies, do have higher specific angular momentum and lower stellar disk surface densities \citep{kim.lee.2013}. In our study we find that the low luminosity dwarf galaxies have the largest $\frac{M_{dyn}}{M(baryon)}$ ratios (Table~2), and the lowest disk surface densities. It is possible that this is due to an oblate halo shape which can be due to larger halo spin. A flattened dark matter distribution would also increase the midplane density of mass in the galaxy. If it has a similar scale height as the gas as assumed in this paper, it can be compared with the mass models derived from rotation curves \citep{carignan.etal.1990}. We will be exploring this in a future study. 

The second origin, the existence of disk dark matter was first postulated by Oort \citep{oort.1932} who used it to explain the vertical equilbrium of old stars in the Galactic disk. It has also been invoked to explain the flaring of the Galactic gas disk \citep{kalberla.etal.2007} and has been included as a component of the Galactic disk in several studies \citep{kramer.randall.2016,fan.etal.2013,saburova.etal.2013}. The origin of disk dark matter has not been extensively explored but studies suggest that it could have formed as a result of mergers with satellite galaxies in $\Lambda$CDM simulations \citep{read.etal.2008}. Other studies suggest that disk dark matter may be distinct from the halo dark matter and is clumpy in nature, because it has settled within the disk \citep{buckley.difranzo.2018}. 

Another important implication of our study is for star formation in the outer disks of galaxies. Star formation is triggered on global scales by processes such as galaxy interactions, mergers, spiral arms and bar instabilities. Or it can be triggered by local events such as gas compression due to supernova shocks, stellar winds and AGN outflows. But on smaller scales it arises due to local disk instabilities that maybe sheared out into small, flocculent spiral arms. In the outer disks of isolated H\,{\sc i} dominated galaxies, there are no clear spiral arms or winds/outflows that can trigger star formation and so local disk instabilties are the main cause for star formation in these regions. Early studies have shown that disk stability is defined by a factor $Q=\frac{\kappa\sigma}{\pi G\Sigma}$ (commonly called the Toomre Q factor), where $\kappa$ is the epicyclic frequency, $\sigma_{s}$is the stellar velocity dispersion and $\Sigma$ is the stellar mass surface density. If $Q<1$, a disk is unstable to small perturbations \citep{safronov.1960,toomre.1964}. An expression for $Q$ has also been derived for disks composed of gas and stars \citep{jog.solomon.1984,rafikov.2001}. In all these cases the disk becomes closer to instability as the mass surface density of baryons increases. 

Many disk galaxies have compact regions of star formation in their outer disks (e.g NGC6946 and NGC628) and are detected in H$\alpha$ emission \citep{ferguson.etal.1998,barnes.etal.2012} or UV emission \citep{thilker.etal.2007,gildepaz.etal.2007,boissier.etal.2007}. However, a purely H\,{\sc i} gas disk may not have the self gravity to become unstable and form stars and the Toomre Q factor due to H\,{\sc i} alone is too high ($Q>1$) \citep{das.etal.2019}. The presence of low luminosity disk mass or halo dark matter associated with the disks can lower the Toomre~Q factor, thus allowing local instabilities to form  and leading to star formation \citep{krumholz.etal.2016}. But not all galaxies may have enough disk mass in their outer regions to allow gravitational instabilities to develop. Hence, only galaxies with enough low luminosity mass or dark matter in their outer disks can support star formation in these extreme environments.

Extended low luminosity or dark matter disks are also important for the inside out growth of galaxy disks driven by gas accreted via galaxy tidal interactions, major mergers, minor mergers \citep{marasco.etal.2019,heald.etal.2011}, or from the intergalactic medium (IGM) from the filaments of the cosmic web 
\citep{noguchi.2018,kleiner.etal.2017,dekel.etal.2009,keres.etal.2005}. Whatever the accretion process, star formation can proceed only if the disk mass surface density $\Sigma$ is high enough to allow local or large scale instabilities to develop. If the outer disk is already close to instability (Q$\sim$1) then any increase in $\Sigma$ due to external gas accretion can further lower the Q value and trigger local star formation. The importance of disk dark matter is that it can increase  $\Sigma$ in the outer extremities and aid the onset of star formation.

\section{Conclusions}

\noindent
{1.~}We have derived an analytical expression for estimating the dark matter in the disks of galaxies using the stellar and H\,{\sc i} gas distributions, and the mean H\,{\sc i} velocity dispersion derived from fitting a single Gaussian function to azimuthally stacked H\,{\sc i} profiles. Our method can be applied to face-on galaxies with extended H\,{\sc i} disks.\\ 
{2.~}We have applied our method to the extended H\,{\sc i} disks of the large disk galaxies NGC628, NGC6946, NGC3184, NGC4214 and the gas rich dwarf galaxies DDO46, DDO63, DDO187. We have used the distribution of H\,{\sc i} velocity dispersion to derive the disk dynamical masses $\Sigma_{dyn}$ and compared it with the disk baryonic masses in the regions $R>R_{25}$ using H\,{\sc i} half disk thickness values of 0.5~kpc for the large galaxies, and half disk thickness values of 0.5~kpc and 1~kpc for the smaller dwarf galaxies. \\
{3.~}Our results show that for the large disk galaxies $\Sigma_{dyn}>\Sigma_{\rm{H\,{\textsc i}}}$ for $R>R_{25}$. But for the dwarf galaxies $\Sigma_{dyn}>\Sigma_{\rm{H\,{\textsc i}}}$ for nearly all radii and is the largest for DDO187.\\
{4.~}Our results mean that either there is a very low luminosity stellar disk in the outer regions of the galaxies, or there is dark matter associated with the outer disks. The latter can occur if the halo is very oblate or if interactions with galaxies lead to the buildup of disk dark matter. The disk dark matter is important for supporting the H\,{\sc i} gas layer in the extreme outer parts of galaxies where there appears to be no stellar disk. \\     
{5.~}The disk dark matter also has important implications for outer disk star formation as $\Sigma_{dm}$ decreases the Toomre Q factor and this can lead to the onset of local disk instabilities, and hence star formation. This is important for explaining the star formation in XUV disk galaxies and H$\alpha$ detected from star formation in the extreme outer disks of some local galaxies. It also suggests that gas accretion in galaxies can lead to outer disk star formation only when the $\Sigma_{dm}$ is large enough to support local disk instabilities in the outer parts of galaxies.

\acknowledgments
The authors gratefully acknowledge IUSSTF grant JC-014/2017 which enabled the author MD to visit CWRU and develop the science presented in this paper. We also thank the anonymous referee for useful comments that improved the paper. 

This research has made use of the NASA/IPAC Extragalactic Database
(NED), which is operated by the Jet Propulsion Laboratory, California Institute of Technology, under contract with
the National Aeronautics and Space Administration. This publication makes use of data products from the Spitzer Survey, which is a joint project of the University of Massachusetts and the Infrared Processing and
Analysis Center/California Institute of Technology, funded by the National Aeronautics and Space Administration
and the National Science Foundation. Funding for the SDSS and SDSS-II has been provided by the Alfred P. Sloan
Foundation, the Participating Institutions, the National Science Foundation, the U.S. Department of Energy, the
National Aeronautics and Space Administration, the Japanese Monbukagakusho, the Max Planck Society, and the
Higher Education Funding Council for England. The SDSS Web Site is http://www.sdss.org/.

The SDSS is managed by the Astrophysical Research Consortium for the Participating Institutions. The Participating
Institutions are the American Museum of Natural History, Astrophysical Institute Potsdam, University of Basel,
University of Cambridge, Case Western Reserve University, University of Chicago, Drexel University, Fermilab, the
Institute for Advanced Study, the Japan Participation Group, Johns Hopkins University, the Joint Institute for Nuclear
Astrophysics, the Kavli Institute for Particle Astrophysics and Cosmology, the Korean Scientist Group, the Chinese
Academy of Sciences (LAMOST), Los Alamos National Laboratory, the Max-Planck-Institute for Astronomy (MPIA),
the Max-Planck-Institute for Astrophysics (MPA), New Mexico State University, Ohio State University, University of
Pittsburgh, University of Portsmouth, Princeton University, the United States Naval Observatory, and the University
of Washington.

We 



\vspace{5mm}
\facilities{SPITZER, VLA}

\software{MIRIAD, SAOIMAGE}




\bibliographystyle{aasjournal}
\bibliography{disk_dm}


\end{document}